\documentclass[10pt, aps,twocolumn,floatfix,showpacs,  nofootinbib,amsmath,amssymb]{revtex4-1}

\usepackage{titlesec}
\usepackage{lipsum} 
\usepackage[a4paper, top=2.2cm, bottom=2.5cm, left=1.8cm, right=1.8cm]{geometry}
\usepackage{graphicx}
\usepackage{graphics}
\usepackage{float}
\usepackage{wrapfig}

\usepackage{blindtext}
\usepackage{setspace}

\usepackage{natbib}

\usepackage{amsmath}
\usepackage{siunitx}

\usepackage[colorlinks=true,linkcolor=blue,citecolor=blue,urlcolor=blue]{hyperref}
\usepackage{graphicx}
\usepackage{tabularx} 
\usepackage{multirow}
\usepackage{lineno}
\usepackage[table]{xcolor}
\definecolor{Gray}{gray}{0.9}
\DeclareSIUnit\ppm{ppm}
\usepackage{makecell}
\usepackage{bibentry}
\DeclareSIUnit{\Mcps}{Mcps}

\newcommand{\biginitial}[1]{\textbf{\LARGE #1}}
\titleformat{\section}
  {\sffamily\normalsize\bfseries}
  {}
  {0pt}
  {\MakeUppercase}

\titleformat{\subsection}
  {\sffamily\normalsize\bfseries}
  {}
  {0pt}
  {}

\titleformat{\subsubsection}
  {\sffamily\normalsize\bfseries}
  {}
  {0pt}
  {}

\titlespacing{\section}
  {0pt}{\baselineskip}{0.2\baselineskip}
\titlespacing{\subsection}
  {0pt}{0.1\baselineskip}{0.1\baselineskip}
\titlespacing{\subsubsection}
  {0pt}{0.1\baselineskip}{0.1\baselineskip}

\newcommand{\reducefigurespace}{\vspace{-10pt}} 
\newcommand{\reducetablespace}{\vspace{-10pt}} 
\newcommand{\reducefigurespaceaftercaption}{\vspace{-1.\baselineskip}}
\newcommand{\ThreeDPi}{3D$\pi$}
\newcommand{\ThreeDPiGammaEnergy}{\SI{511}{\kilo\eV}}
\newcommand{\grs}{\mbox{$\gamma$-rays}}
\newcommand{\gr}{\mbox{$\gamma$-ray}}
\newcommand{\ArDimerSingletMeanLife}{\SI{6}{\nano\second}}
\newcommand{\LYSOTimeConstant}{\SI{42}{\nano\second}}
\newcommand{\LArNormalTemperature}{\SI{87}{\kelvin}}
\newcommand{\ArWaveLength}{\SI{128}{\nano\meter}}
\newcommand{\XeWaveLength}{\SI{172}{\nano\meter}}
\newcommand{\TPBWaveLength}{\SI{420}{\nano\meter}}

\newcommand{\SigmaSiPM}
{\mbox{$\sigma_{\text{\scriptsize SiPM}}$}}
\newcommand{\CommercialPETBestTimingRes}{\SI{100}{\pico\second}}

\newcommand{\ThreeDPiFOV}{\SI{200}{\centi\meter}}
\newcommand{\ThreeDPiID}{\SI{90}{\centi\meter}}
\newcommand{\ThreeDPiOD}{\SI{64}{\centi\meter}}
\newcommand{\ThreeDPiRingNumber}{\num{9}}
\newcommand{\ThreeDPiLArLayerThickness}{\SI{2.11}{\centi\meter}}
\newcommand{\ThreeDPiPTFEThickness}{\SI{0.75}{\mm}}
\newcommand{\ThreeDPiSiPMSize}{\qtyproduct[product-units = power]{10 x 10}{\square\mm}}
\newcommand{\ThreeDPiLArThickness}{\SI{18}{\milli\meter}}
\newcommand{\ThreeDPiNumSiPM}{\num{1E6}}
\newcommand{\Geant}{\mbox{Geant4}}
\newcommand{\SigmaPar}{\mbox{$\sigma_\parallel$}}
\newcommand{\SigmaPerp}{\mbox{$\sigma_\perp$}}

\begin{document}

\makeatletter
\renewcommand{\maketitle}{%
  \bgroup
  \setlength{\parindent}{0pt} 
  \begin{flushleft}
    \bfseries{\@title} \\
    {\normalfont\normalsize{\@author}}
  \end{flushleft}
  \egroup
}

\title{3D$\pi$: Three-Dimensional Positron Imaging, A Novel Total-Body PET Scanner Using Xenon-Doped Liquid Argon Scintillator}
\normalfont\normalsize\author{Azam Zabihi$^{1}$\thanks{Email: mail@domain.com}, Xinran Li$^{2}$, Alejandro Ramirez$^{3}$, Manuel D. Da Rocha Rolo$^{4}$, Davide Franco$^{5}$, Federico Gabriele$^{6}$, Cristiano Galbiati$^{7,8}$, Michela Lai$^{9,10}$, Daniel R. Marlow $^{7}$, Andrew Renshaw$^{3}$, Shawn Westerdale$^{9}$, Masayuki Wada$^{1,10}$\\
  $^{1}$AstroCeNT, Nicolaus Copernicus Astronomical Center of the Polish Academy of Sciences, Warsaw, Poland, $^{2}$Lawrence Berkeley National Laboratory, Berkeley, CA, USA, $^{3}$Department of Physics, University of Houston, Houston, TX, USA,  $^{4}$INFN Torino, Torino, Italy, $^{5}$APC, Universit\'e de Paris, CNRS, Astroparticule et Cosmologie, Paris, France, $^{6}$INFN Cagliari, Cagliari, Italy, $^{7}$Physics Department, Princeton University, Princeton, NJ, USA, $^{8}$ Gran Sasso Science Institute, L’Aquila, Italy, $^{9}$Department of Physics and Astronomy, University of California, Riverside, CA, USA, $^{10}$Physics Department, Universit\`a degli Studi di Cagliari, Cagliari, Italy. \\
  }

\onecolumngrid
\maketitle
\twocolumngrid
\clearpage


\fontsize{9pt}{11pt}\selectfont
\noindent\rule{\columnwidth}{1.2pt}\vspace{-7pt}
\section{}
\vspace{-\baselineskip}
\label{sec:Abstract}
\reducefigurespace
{\bf Objective:} This paper introduces a novel PET imaging methodology called 3-dimensional positron imaging (\ThreeDPi), which integrates total-body (TB) coverage, time-of-flight (TOF) technology, ultra-low dose imaging capabilities, and ultra-fast readout electronics inspired by emerging technology from the DarkSide collaboration.
{\bf Approach:} The study evaluates the performance of \ThreeDPi\ using Monte Carlo simulations based on NEMA NU 2-2018 protocols. The methodology employs a homogenous, monolithic scintillator composed of liquid argon (LAr) doped with xenon (Xe) with silicon photomultipliers (SiPM) operating at cryogenic temperatures. {\bf Main results:} Significant enhancements in system performance are observed, with the \ThreeDPi\ system achieving a noise equivalent count rate (NECR) of \SI{3.2}{\Mcps} which is approximately two times higher than uEXPLORER's peak NECR (\SI{1.5}{\Mcps}) at \SI{17.3}(kBq/mL). Spatial resolution measurements show an average FWHM of \SI{2.7}{\mm} across both axial positions. The system exhibits superior sensitivity, with values reaching \SI{373}{kcps/MBq} with a line source at the center of the field of view. Additionally, \ThreeDPi\ achieves a TOF resolution of \SI{151}{\ps} at \SI{5.3}{kBq/mL}, highlighting its potential to produce high-quality images with reduced noise levels.
{\bf Significance:} The study underscores the potential of \ThreeDPi\ in improving PET imaging performance, offering the potential for shorter scan times and reduced radiation exposure for patients. The Xe-doped LAr offers advantages such as fast scintillation, enhanced light yield, and cost-effectiveness. Future research will focus on optimizing system geometry and further refining reconstruction algorithms to exploit the strengths of \ThreeDPi\ for clinical applications.

{\bf Key Words:} TOF-PET, \ThreeDPi, Liquid Argon, total-body imaging, performance evaluation;

\noindent\rule{\columnwidth}{1.2pt}\vspace{-7pt}
\section{}
\label{sec:Intro}
\reducefigurespace
\biginitial{T}his paper presents a shift from traditional modular-based scintillators positron emission tomography (PET) to a homogenous, monolithic scintillator for design readout by multiple SiPMs simultaneously. The design forms the foundation of \ThreeDPi, a novel PET methodology leveraging TB coverage, TOF technology, ultra-low dose imaging capabilities, and ultra-fast readout electronics for significantly reduced scan times. Our work is inspired by emerging technology from direct dark matter searches, particularly the successful methodologies developed within the DarkSide collaboration to advance cryogenic photosensor, LAr detector technology, and low-radioactivity argon procurement~\cite{Agnes:2016fz, 8113575, Aalseth:2017hu}. 
\subsection{The Principle of \ThreeDPi}
\begin{figure}[t]
\begin{center}
\includegraphics[width=0.85\columnwidth]{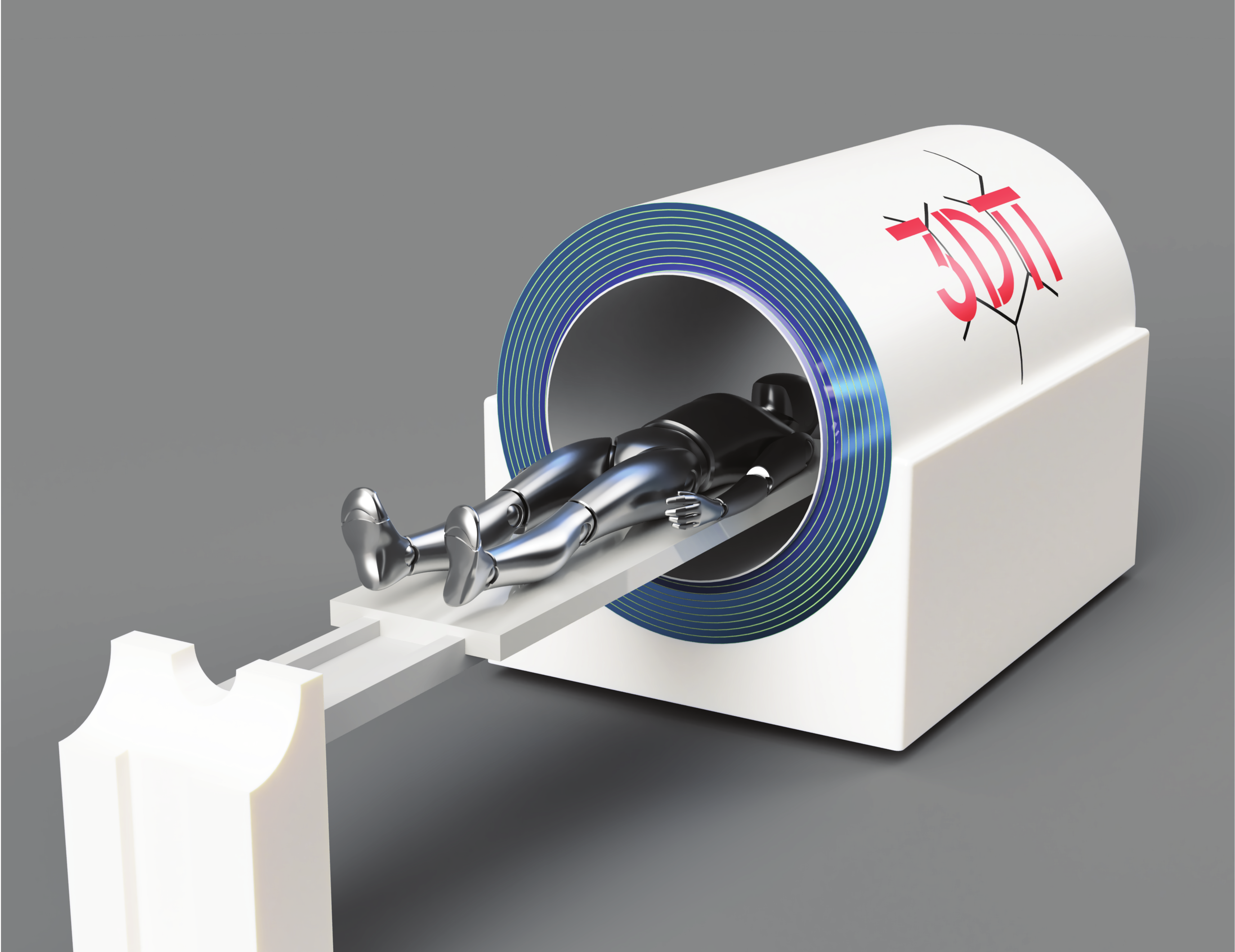}
\end{center}
\reducefigurespace
\caption{CAD model of the \ThreeDPi\ detector with a human phantom. The ends of the cylinder are left open without a full cryostatic enclosure to show the LAr layers.}
\label{fig:CADModel}
\reducefigurespace
\reducefigurespace
\end{figure}

The \ThreeDPi\ detector features a monolithic barrel filled with Xe-doped LAr, viewed by several cylindrical layers of fast cryogenic SiPMs as schematized in Fig.~\ref{fig:CADModel}. The main design concepts are discussed below.

\subsubsection{LAr as a Monolithic Scintillator}
\label{ssec:LAr}
The excellent scintillation properties of LAr make it a prime candidate for large homogeneous scintillation volumes for TOF-PET application. LAr has a short scintillation time constant of \ArDimerSingletMeanLife~\cite{Kubota:1978bi}, approximately \num{7} times faster than the \LYSOTimeConstant\ decay time for LYSO crystals~\cite{Mao_2013}. LAr induces \num{40000} photons per \si{\MeV} of deposited energy~\cite{DOKE1990617}, \SI{25}{\percent} higher than the LYSO crystal yield~\cite{Mao_2013}. As a liquid that is transparent to its own scintillation light, LAr can be scaled to very large volumes without a significant cost increase. The continuous nature and absence of small-cell segmentation reduce dead spaces and improve the reconstruction of single Compton scatters, enhancing imaging precision. The cryogenic temperature (\LArNormalTemperature) of LAr enables efficient cooling of SiPMs and their front-end electronics, ensuring stable temperature conditions and consistent detector response. As an option, low-radioactivity argon~\cite{Agnes:2016fz} can be used to mitigate the rate of random coincidence events.

\subsubsection{Xenon Doping for Enhanced Performance}
\label{ssec:XenonDoping }
From the \ThreeDPiGammaEnergy\ \grs\ only about \SI{30}{\percent} of the scintillation photons are emitted with the short decay time, the remaining \SI{70}{\percent} produce scintillation photons even after \SI{1}{\micro\second} in pure LAr~\cite{Kubota:1978bi}.
Considering that \grs\ may scatter multiple times in the LAr, this long decay time introduces the possibility of photons detected from a previous event overlapping the current event. This results in degrading the Signal to Noise Ratio (SNR) and thus image quality. It would be highly beneficial to suppress this long-lifetime component of the LAr scintillation to avoid limitations on tracer dosage.

Additionally, LAr scintillates at \ArWaveLength, deep in the vacuum-ultraviolet (VUV) and even modern photosensors have very low efficiency for detecting VUV photons, typically less than \SI{15}{\percent}~\cite{Pershing_2022}. It is therefore customary to shift the \ArWaveLength\ scintillation photons using an organic wavelength shifter (WLS) such as tetraphenylbutadiene (TPB). However, the absorption and reemission time of TPB is on the order of a \si{\nano\second}~\cite{Flournoy:1994bx}, which will degrade the overall timing resolution and cut into the gain achieved by moving into a faster scintillation readout. Doping the scintillator with xenon offers a potential solution. 
Suppression of the long-lifetime component of LAr scintillation light is observed with xenon doping~\cite{KUBOTA199371}. This enhancement allows the scanner to handle higher decay rates, potentially accommodating increased patient dosages as needed for specific applications. Doping LAr with xeon causes scintillation light to be emitted at longer wavelengths, \XeWaveLength\ rather than  \ArWaveLength~\cite{Cheshnovsky1972}, which results in \SI{\sim 70}{\percent} higher light detection efficiency than the VUV sensitive  SiPMs~\cite{Ootani:2015ga, Pershing_2022}. Xenon doping presents an economical advantage compared to pure xenon, historically used in PET applications~\cite{lavoie1976liquid, ROMOLUQUE2020162397}.

For a comprehensive review of early and recent studies on the scintillation of Xe-doped argon and its thermodynamic stability, please refer to~\cite{Galbiati_2021, KUBOTA199371, PhysRevC108045503}. These studies have shown stable LAr+Xe mixture up to \SI{2}{\percent} in molar fraction and shortening of the slow decay constant to \SI{100}{\nano\second}. 

\subsubsection{Fast Cryogenic SiPMs}
\label{ssec:LightDetection}
Some of the recent SiPMs have an outstanding timing resolution (\SigmaSiPM), already below \CommercialPETBestTimingRes\ and still improving~\cite{Gundacker_2020}. The photon detection efficiency (PDE) is \SI{\sim 40}{\percent}~\cite{Acerbi:2015ca} at \TPBWaveLength, \SI{\sim 20}{\percent} at \XeWaveLength~\cite{Ootani:2015ga} and \SI{\sim 12}{\percent} at \ArWaveLength~\cite{Pershing_2022}. Because the Xe-doping studies mentioned above used photomultiplier tubes (PMT) as the photosensors, which are not directly sensitive to the \XeWaveLength\ light, it is expected that using SiPMs with better sensitivity to this wavelength will further enhance the potential light yield. This increase in light yield improves the resolution of the Compton scattering vertex \gr\ in the LAr, leading to an increased resolution of the positron annihilation vertex, thus improving image quality. Recent developments in advanced noise suppression techniques at cryogenic temperature indicate the possibility that SiPMs could run at high over-voltages to achieve a very high PDE~\cite{Acerbi:2017gy}. Further increases in PDE can be obtained by using microlenses found in CMOS imaging detectors~\cite{Fossum:1997jp}.

Operating the LAr+Xe mixture at the boiling point of LAr results in a remarkable 100-fold reduction in the dark count rate (DCR) of SiPM compared to DCR at liquid xenon temperature as shown in Fig.~6 (left) in Ref.~\cite{Aalseth:2017hu}. This reduction not only enhances the timing capabilities of the devices but also improves the SNR.

This study aims to comprehensively evaluate the efficacy and performance of the \ThreeDPi\ methodology in modern PET imaging, focusing on its geometry, reconstruction methods, and a comparative analysis against established PET technology benchmarks.

\section{Materials and methods}
\label{sec:MandM}
\subsection{\ThreeDPi\ Geometry}
\label{ssec:Geometry}
\begin{figure}[tbh!]
 \reducefigurespace
\begin{center}
\includegraphics[width=\columnwidth]{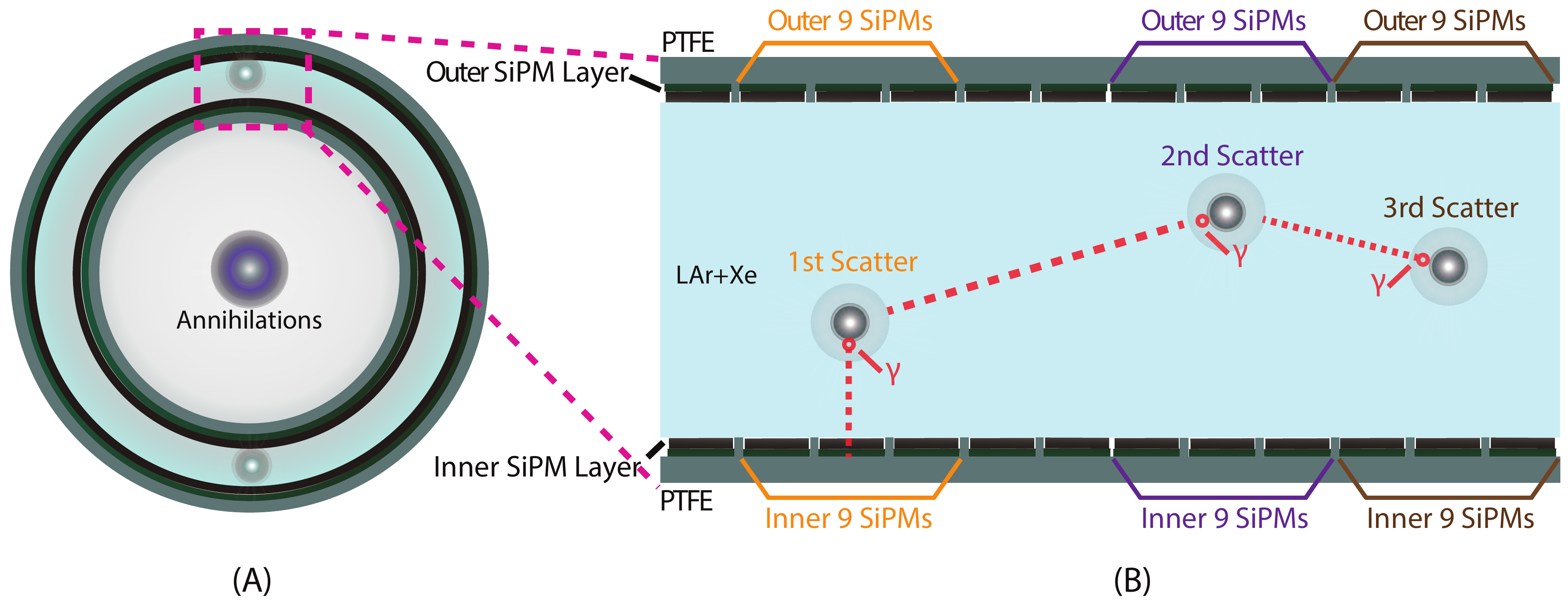}
\end{center}
 \reducefigurespace
\caption{{\bf A:} An illustration of a single detection layer of the \ThreeDPi\ detector with the LAr+Xe scintillation configuration. {\bf B:} Each detection layer contains both outer and inner layers of PTFE supporting material with an array of SiPMs and visualization of SiPM arrangement for Compton scatter position calculation.}
\label{fig:DetectorGammaScatter}
 \reducefigurespace
\end{figure}

We conduct Monte Carlo simulations to assess the feasibility of constructing the \ThreeDPi\ detector, a TB-TOF-PET system with annular cylinder geometry, as shown in Fig.~\ref{fig:CADModel} and Fig.~\ref{fig:DetectorGammaScatter}. The annular cylinder, filled with LAr, has a bore diameter of \ThreeDPiID, a transaxial field of view (TFOV) of \ThreeDPiOD, and an axial FOV (AFOV) of \ThreeDPiFOV. The annular cylinder is subdivided into \ThreeDPiRingNumber\ different annular, concentric cylinders, each with a radial thickness \ThreeDPiLArLayerThickness\ and a supporting \ThreeDPiPTFEThickness\ thick PTFE layer. Each smaller annular cylinder is completely covered with 
\ThreeDPiSiPMSize\ 
SiPM pixels, covering the inner and outer surfaces of the scintillator. The choice of annular cylinder geometry offers several advantages. It provides a uniform and scalable structure conducive to efficient photon detection and data acquisition. Additionally, the concentric arrangement of layers facilitates precise spatial localization of gamma interactions, thereby enhancing sensitivity and spatial resolution. Furthermore, this geometry enables the exploration of novel reconstruction algorithms tailored to the unique characteristics of \ThreeDPi. The main parameters of the \ThreeDPi\ geometry are summarized in Table.~\ref{tab:Parameters}. 
\begin{table}[htb!]
\vspace{-6pt}
\centering
\caption{\ThreeDPi\ detector (\Geant\ geometry) parameters.}
\label{tab:Parameters}
\begin{tabularx}{\columnwidth}{X r}
\hline
Parameter				&Value\\
\Xhline{2pt}
Bore diameter			&\ThreeDPiID\ \\
Transaxial FOV			&\ThreeDPiOD\ \\
Axial FOV				&\ThreeDPiFOV\ \\
LAr thickness per layer			&\ThreeDPiLArThickness\ \\
Number of LAr layers	&\ThreeDPiRingNumber\ \\
SiPM size			&\ThreeDPiSiPMSize\ \\
Number of SiPMs 		&\ThreeDPiNumSiPM\ \\
\hline
\end{tabularx}
\vspace{-10pt}
\end{table}
For simplicity in the simulations, the front-end electronics layers for the SiPMs were omitted. In between the SiPMs and the PTFE there is an \SI{18}{\mm} thick region of LAr. The total thickness of the LAr volume in the radial direction is \SI{16.2}{\cm}. A \SI{6}{\mm} thick titanium shell is also included just inside the inner radius of the LAr, accounting for the minimal thickness required for the cryostat that would contain the LAr. Our primary configuration is doping the LAr with xenon to achieve a minimal concentration of \SI{0.5}{\percent} by molar fraction without using TPB (referred to as LAr+Xe). This concentration level is essential for fast wavelength shifting of scintillation light, as discussed previously. We also introduce an additional configuration using pure LAr with all surfaces coated with TPB (referred to as LAr+TPB) to explore the advantages of doping with xenon. We utilized a Monte Carlo simulation package, adapted from the DarkSide dark matter search experiment~\cite{Agnes:2017cz} based on the \Geant\ toolkit~\cite{Agostinelli:2003fg}. This package tracks charged particles, \grs, Cherenkov photons and the scintillation photons, enabling full reconstruction of all simulated events. The \ThreeDPi\ geometry has been incorporated into the simulation package and was used for all simulations, with variations in source geometries and phantoms for specific measurements.

\subsection{Reconstruction Methods}
\subsubsection{Compton Scatter Reconstruction}
\label{sec:ComScatterRec}
To reconstruct the position of \gr\ Compton scatters, we first identify peaks in the photon hit counts on the SiPM location map. For each peak (representing a Compton scatter), the position along z and azimuthal axes are estimated using a hit-weighted average of the SiPM with the peak and the adjacent SiPMs (see Fig.~\ref{fig:DetectorGammaScatter}). The depth of the Compton scatter in the radial direction is then calculated from the asymmetry in the photon counts detected by the inner and outer two sets of nine SiPMs surrounding the peak. As illustrated in Supplemental Fig.~\ref{fig:Asymmetry} (In this paper, figures, equations, and tables labeled with `S' represent supplemental materials), the relationship between the radial position and the asymmetry is linear. 
\subsubsection{Annihilation Vertex Reconstruction}
\label{sec:AnniRec}
The Compton scatters of the same \gr\ are clustered based on space-time separation criteria. Then, we define each event as back-to-back \gr\ pairs within a coincidence window of \SI{4}{ns} based on the times and positions of each \gr's first scatters. The annihilation vertex position is determined by TOF along the line of response (LOR) defined by these two earliest scatters.

\subsection{Energy Cut}
\label{ssec:OPHcut}
Similar to conventional PET scanners that utilize energy windows to mitigate scattering events, we introduce optical photon cut (OPC) for the same purpose. The OPC is implemented by setting a threshold on the number of detected optical photons of all Compton scatters in each gamma, offering an alternative method for filtering out scattered coincident events, which are scattered in patients' bodies or the detector's dead volume before reaching the scintillator volume. Events from \grs\ that produce fewer than \num{3000} detected photons are rejected. This threshold is determined from the simulation, which will be discussed in the \hyperref[sec:Results]{Result} section. We consistently apply this OPC threshold throughout our assessments unless otherwise specified.

\subsubsection{Image Reconstruction}
\label{ssec:IRM-FBP} 
\begin{figure}[b]
 \vspace{-18pt}
\begin{center}
\includegraphics[width=0.35\columnwidth]{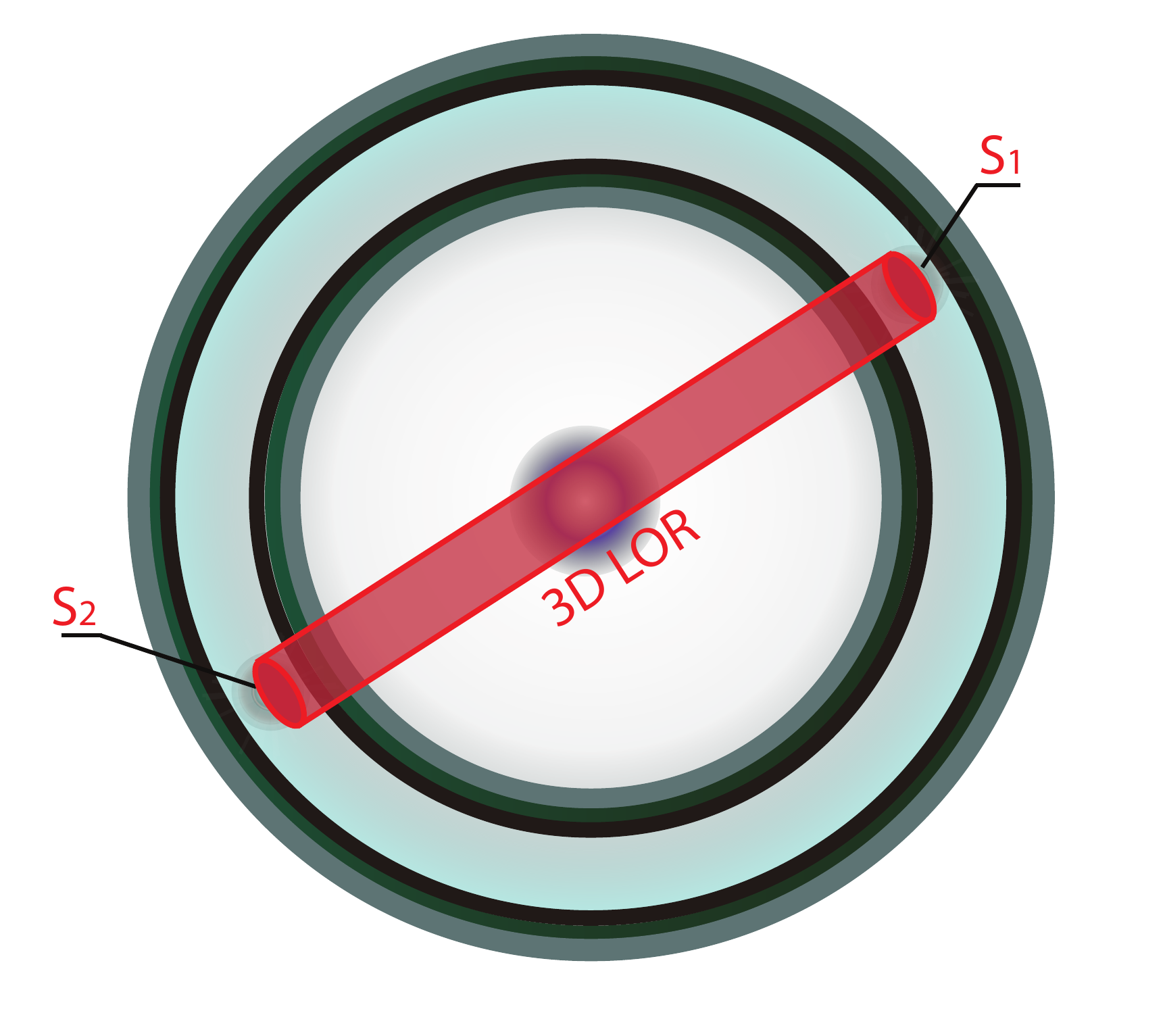}
\includegraphics[width=0.35\columnwidth]{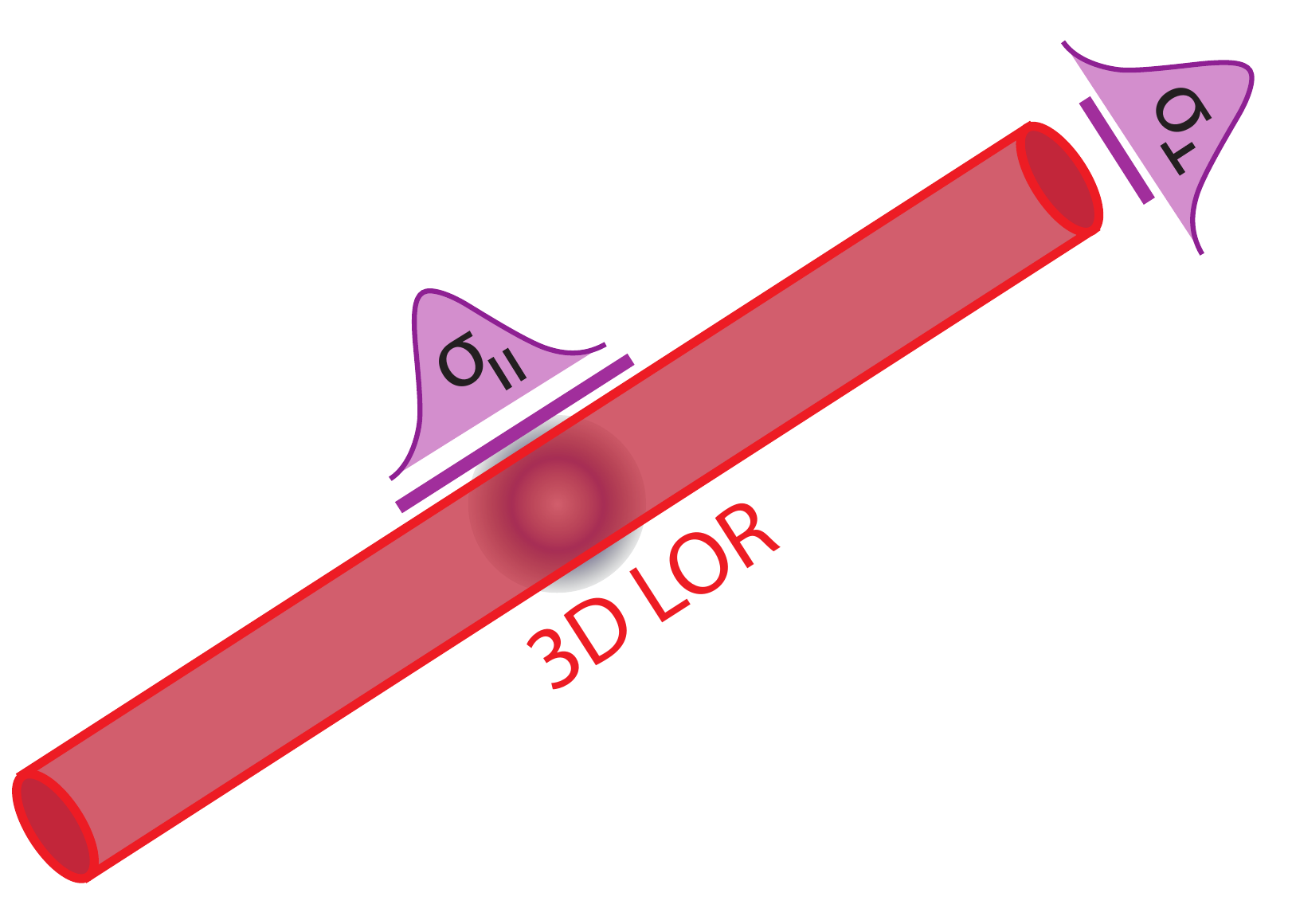}
\end{center}
 \reducefigurespace
\caption{Resolutions parallel and perpendicular to the LOR.}
\label{fig:smearedLOR}
 \vspace{-18pt}
\end{figure}

While the development of new, faster, and more robust algorithms shows promise, in this work we use a straightforward filtered deconvolution reconstruction (FDR) approach for its simplicity and ease of comparison. This method involves a few steps. 
First, each reconstructed annihilation point is filled to a 3D histogram as a probability distribution function of 3D Gaussian aligning the long principal axis to the LOR with a mean at the annihilation point and \SigmaPar\ and \SigmaPerp\ capturing detector resolutions parallel and perpendicular to the LOR of the event (see Fig.~\ref{fig:smearedLOR}), respectively. Those resolutions are extracted from the simulation with point sources as discussed in the \hyperref[sec:Results]{Result} section. 
After filling all events, the data is transformed into the Fourier domain using a Fast Fourier Transform to deconvolve the detector resolutions with a filter of
\begin{equation}
\frac{k}{2\pi^{2}\mathrm{erf}(k\SigmaPar/\sqrt{2})}, \tag*{Eq. 1}
\label{eq:FDReq1}
\end{equation}
where $k$ is the vector in frequency space and erf(x) is the error function.
The filter is a reciprocal of the 3D Fourier transformed resolution function $\exp(-r^2/2\SigmaPar^2)/r^2$. As we see later since \SigmaPerp\ contribution is negligible compared to \SigmaPar, only \SigmaPar\ is used to represent the detector resolution. 
After inverse Fourier transform, the image $f(r)$ derived from the positron annihilation probability density plot $f'(r)$ will be:
\begin{equation}
\setlength{\abovedisplayskip}{6pt} 
\setlength{\belowdisplayskip}{6pt} 
f(r) = \mathcal{F}^{-1}\left(\mathcal{F}(f'(r))\frac{k}{2\pi^{2}\mathrm{erf}(k\SigmaPar/\sqrt{2})}\right). \tag*{Eq. 2}
\label{eq:FDReq2}
\end{equation}

This method not only integrates spatial information through event weighting but also takes advantage of TOF data via the custom filter in the frequency domain. This synergy between spatial and temporal cues results in sharper images with improved resolution. Validation of the FDR method is discussed through~\hyperref[sec:Results]{Results} section.
 
\subsection{Simulation Parameters for SiPM Implementation}
In this work, the PDE for light above \SI{300}{\nm} is taken from near-UV sensitive, high-density SiPMs (NUV-HD), as depicted in Fig.~3 of Ref.~\cite{gola2019nuv}, measured with an over-voltage of \SI{4}{V}. The NUV-HD SiPMs are not sensitive to light below \SI{300}{\nm} due to an anti-reflection coating not optimized for those wavelengths. We assume better optimization in this aspect in our scanner and PDE values of \SI{30}{\percent} between \SI{300}{\nm} and \SI{130}{\nm}, and \SI{16}{\percent} between \SI{130}{\nm} and \SI{120}{\nm}~\cite{Pershing_2022}, with negligible efficiency below \SI{120}{\nm}. Those PDEs are implemented in our simulation with consideration of the surface reflectivity measured in Ref.~\cite{Lv_2020}. Moreover, a \SigmaSiPM\ of \SI{60}{\ps} is assumed. 

\subsection{NEMA NU 2-2018 Tests}
For direct comparison with commercially available PET scanners, we rigorously followed the testing guidelines established in the NU 2-2018 National Electrical Manufacturers Association (NEMA) document~\cite{national2018nu}. We used $^{18}$F as the radioisotope for all conducted tests.
\subsubsection{Spatial Resolution}
\label{ssec:SpatialRes-NEMA}
Six point sources, each representing a small volume of concentrated radioactivity, were simulated at various radial positions, positioned at the center and offset of the AFOV (Fig.~\ref{fig:NEMA-Spatial-Res-Geometry}). Spatial resolution was evaluated based on the distribution of the reconstructed annihilation points using FDR. Each position was acquired with a minimum of 125,000 counts. 
\subsubsection{Count Rate Performance}
\label{ssec:Scatter}
Details of the phantom and source used in the simulation are provided in Fig.~\ref{fig:SFPh} A and B. Total, true, and scattered plus random coincidence events were computed according to the specifications outlined in Fig.~4-2 of the NEMA standards. Basically, the scattered plus random coincidence is estimated based on the events reconstructed away from the line source. The true coincidence is defined as the events reconstructed close to the line source after subtracting the estimated contribution from the scattered plus random coincidence in the vicinity of the line source. Subsequently, random events were determined based on the Monte Carlo simulation information and separated from scattered events. Those categorizations are done without FDR. Coincidence event count rates were recorded across a range of activity concentrations from low (\SI{0.1}{kBq/mL}) to high activity concentration (\SI{40}{kBq/mL}) to determine the NECR and scatter fraction (SF) using data acquired from 500,000 prompt events. NECR is a metric that quantifies the scanner's ability to resolve true coincident events and is defined as
\begin{equation}
\setlength{\abovedisplayskip}{6pt} 
\setlength{\belowdisplayskip}{6pt} 
\small 
\textrm{NECR} = \frac{R_{t}^2}{R_{tot}}, \tag*{Eq. 3}
\label{eq:NECReq}
\end{equation}
where $R_{t}$ represents the rate of true coincident events, and $R_{tot}$ denotes the total event rate. SF is the ratio of scattered coincident events to total events.
\subsubsection{Sensitivity}
\label{ssec:Sensitivity}
Figure~\ref{fig:SPh} A and B provide details of the phantom utilized in the simulation. The phantom consists of five concentric sleeves surrounding a water cylinder. Within the cylinder, there is a \SI{2}{MBq} $^{18}$F line source. The test aims to measure the rate of collected gammas in the limit of no absorption layer, or in this case, sleeve. To obtain an attenuation-free measurement, successive measurements are conducted using a uniform line source surrounded by known absorbers. The measurement was performed at the origin of the TFOV and at a radial offset of 10 cm from the center, to test the sensitivity at different radial positions without FDR. Additionally, the axial sensitivity profile is provided for the smallest tube at the center of the TFOV.
\subsubsection{Image Quality}
\label{ssec:Image}
The phantom consists of four components: a body phantom, six spheres, a cylindrical insert to simulate lung attenuation, and a test phantom (Fig.~\ref{fig:IQPh} A, B, and C). To meet the NEMA NU 2-2018 guidelines, we filled the body phantom, as a background activity, with 5.3 kBq/mL of $^{18}$F. The six spheres were filled with a concentration of four times that of the background activity. The line source of the test phantom was filled with an activity concentration equal to the background activity concentration level. The cylindrical insert was filled with a low atomic number material, Styrofoam in our case, with an average density of \SI{0.30}{\gram/\milli\liter} and centered inside the body phantom to simulate the attenuation of the lung. The phantom was positioned on a patient table and adjusted so that the lung insert was centered in the TFOV. 

Image reconstruction was performed using the FDR method. Various metrics were computed in accordance with the NEMA NU 2-2018 guidelines to assess image quality. These metrics include the percent contrast (PC) for each hot sphere, the percent background variability (BV) for different sizes of background regions of interest (ROI), and the average relative error (RE) in the lung region (Eq.~\ref{eq:PC},~\ref{eq:BV}, and~\ref{eq:RE}). 
\subsubsection{TOF Resolution}
\label{ssec:TOF}
The source distribution and the phantom are the same as those described in the \hyperref[ssec:Scatter]{Count Rate Performance} section. Assumed the closest point on the LOR to the line source as a true point, the full width at half maximum (FWHM) of the time difference distribution between the measured TOF and TOF from the true point is reported as TOF resolution. Similar to the \hyperref[ssec:Scatter]{Count Rate Performance} section, we removed the scatter and random contributions as described in the NEMA guidelines.

\section{Results}
\label{sec:Results}
\subsubsection{Resolution Dependence on SiPM Parameters} 
To optimize and improve the performance, it is important to understand how the resolutions are affected by the detector parameters, such as PDE and time resolution of SiPMs. We study the resolution of our scanner by evaluating both \SigmaPar\ and \SigmaPerp\ to the LOR (Fig.~\ref{fig:smearedLOR}) without FDR.
For the reconstruction of the annihilation positions in the direction parallel to the LOR \SigmaPar, there are two sources of fluctuation.
The first is timing fluctuation due to the time resolution of the SiPMs and their electronics, given by \SigmaSiPM, and the statistics of detected photons related to the PDE of SiPMs. 
The second is the uncertainty due to the imperfect position reconstruction within the annular cylinder (Fig.~\ref{fig:Asymmetry}). The dominant of these two is the former timing uncertainty.
The lattice size of the SiPM layout is the ultimate limitation to the resolution on the perpendicular plane to the LOR \SigmaPerp. It has a minor contribution to the total uncertainty. 

Those two resolutions, \SigmaPar\ and \SigmaPerp, are assessed based on the reconstructed positions of point sources for both the LAr+TPB and LAr+Xe configurations as functions of \SigmaSiPM\ and PDE. The OPC was not applied during this evaluation because, in some cases with low PDE, there were not enough detected photons compared to the OPC threshold of \num{3000}. The results are provided in \ref{fig:Xe-SigmasVsSiPM} and Fig.~\ref{fig:TPB-SigmasVsSiPM}, illustrating changes in resolutions corresponding to parameter variations.
In summary, the contribution of \SigmaPerp\ is negligible compared to \SigmaPar\ and does not depend on the SiPM parameters for both configurations. In the LAr+Xe configuration, \SigmaPar\ increases exponentially as the PDE decreases below \SI{30}{\percent} and for PDE above \SI{30}{\percent}, \SigmaPar\ improves linearly. The \SigmaSiPM\ dependency of \SigmaPar\ is linear; \SI{\sim 1}{\cm} to \SI{\sim 2.3}{\cm} as \SigmaSiPM\ changes from \SI{20}{ps} to \SI{160}{ps} with a fixed PDE at \SI{30}{\percent}.
Thus, it is important to have at least \SI{30}{\percent} PDE to gain the advantage from the good time resolution of SiPMs. In the LAr+TPB configuration, the dependencies are similar but the resolutions are worse by \SI{\sim1}{cm}.
The spatial resolutions with the specifications achievable with current SiPM technologies~\cite{Ootani:2015ga} are summarized in Table~\ref{tab:Xe-TPBParameters}. 
\begin{table}[H]
\reducetablespace
\centering
\caption{Summary of spatial resolution parallel and perpendicular to LOR and SiPM parameters without OPC and FDR.}
\label{tab:Xe-TPBParameters}
\begin{tabularx}{\columnwidth}{X X X |  >{\centering\arraybackslash}X >{\centering\arraybackslash}X}
\hline
      & PDE &   \SigmaSiPM\    & \SigmaPar\  & \SigmaPerp\ \\
\Xhline{2pt}
LAr+TPB  & 40$\%$ &  60 ps & 27.2 mm & 1 mm\\
LAr+Xe  & 25$\%$ & 60 ps  & 15.1 mm  & 1 mm\\
\hline
\end{tabularx}
\vspace{-10pt}
\end{table}
Notably, LAr+Xe demonstrates a factor of \num{1.8} improvement in resolution compared to LAr+TPB, attributed to enhanced fast decay components of scintillation light and direct sensitivity of SiPM to \XeWaveLength\ scintillation light.
\subsection{Spatial Resolution}
\begin{figure}[h]
\reducefigurespace
\begin{center}
\includegraphics[width=0.45\columnwidth]{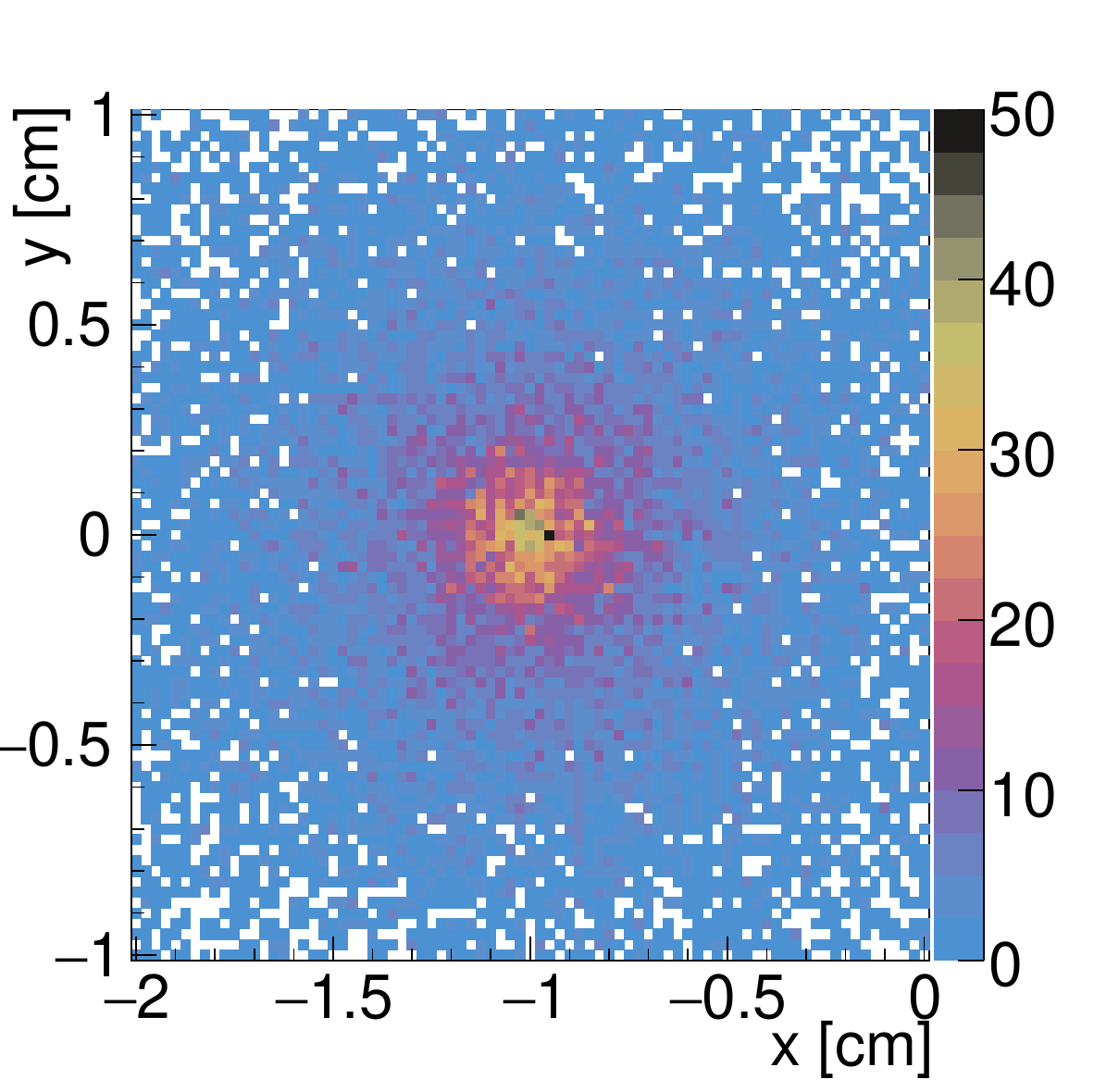}
\includegraphics[width=0.45\columnwidth]{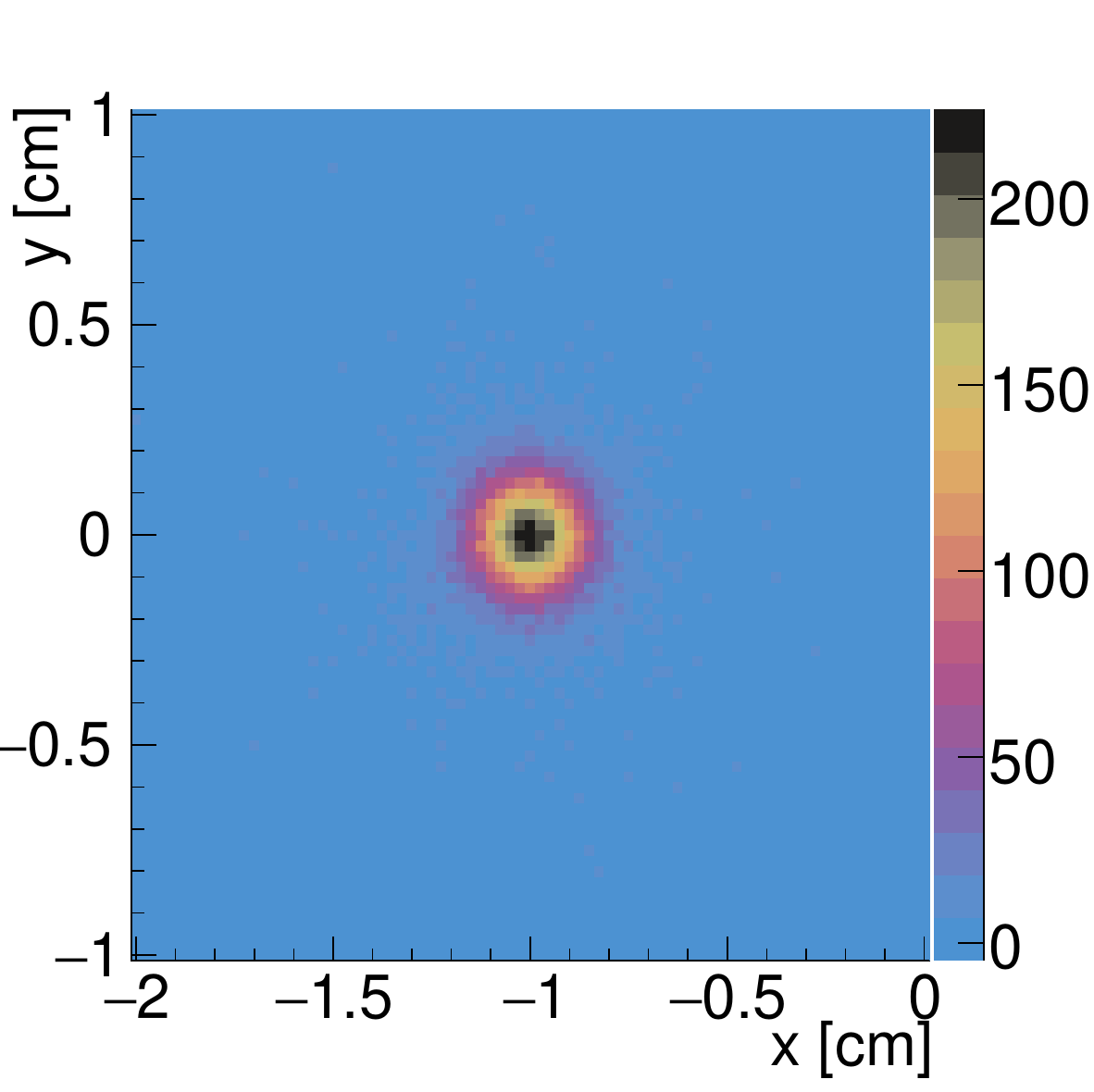}
\end{center}
\reducefigurespace
\reducefigurespace
\caption{{\bf Left:} Raw Data and {\bf Right:} FDR Image, {\bf Configuration:} LAr+Xe and for a point sources at 1 cm radial offset and at the center of of the AFOV.}
\label{fig:RAW-FDR}
\reducefigurespace
\end{figure}

\begin{table*}[t]
\centering
\caption{ Radial, tangential, and axial resolutions for both configurations and compared with Explorer~\cite{spencer2021performance}.}
\label{tab:SpatialResolution}
\begin{tabularx}{1.99\columnwidth}{X  >{\centering}X  >{\centering}X  >{\centering}X  >{\centering}X  >{\centering}X  >{\centering}X >{\raggedleft\arraybackslash}X}
\hline
\multicolumn{3}{c}{Source position}  & \multicolumn{2}{c}{LAr+Xe} & \multicolumn{2}{c}{LAr+TPB}  & uExplorer \\ \cline{4-8}
\multicolumn{3}{c}{[cm]}& FWHM \tiny [mm] &  FWTM \tiny [mm] &  FWHM \tiny [mm]  &  FWTM \tiny [mm]  & FWHM \tiny [mm] \\ 
 \Xhline{2pt}
\multirow{9}{*}{1/2 AFOV} &\multirow{3}{*}{Radial}& 1 & 2.8 & 6.6 & 2.8 & 6.6 &3.0 \\
& & 10 & 2.8 & 7.8 & 2.7 & 5.8 & 3.4\\ 
& & 20 & 2.6 & 7.1 & 2.8 & 5.9 & 4.7  \\ \cline{2-8}
&\multirow{3}{*}{Tangential}& 1 & 2.7 & 7.8 & 2.9 & 8.0 & 3.0 \\
& & 10 & 2.8 & 7.8 & 2.6 & 6.7 & 3.1\\ 
& & 20 & 2.9 & 7.9 & 2.9 & 7.0 & 4.0 \\\cline{2-8}
& \multirow{3}{*}{Axial}& 1 & 2.5 & 5.2 & 2.5 & 5.2 & 2.8 \\
& & 10 & 2.6 & 5.5 & 2.6 & 5.5 & 3.2\\ 
& & 20 & 2.5 & 5.2 & 2.7 & 5.5 & 3.2  \\  \hline 
\multirow{9}{*}{1/8 AFOV} &\multirow{3}{*}{Radial}& 1 & 3.1 & 9.1 & 2.5 & 5.6 & 3.0 \\
& & 10 & 2.8 & 8.2 & 2.8 & 6.1 & 3.6\\ 
& & 20 & 2.8 & 7.2 & 2.9 & 5.9 & 4.6  \\ \cline{2-8}
&\multirow{3}{*}{Tangential}& 1 & 3.1 & 9.0 & 2.7 & 6.8 & 2.9 \\
& & 10 & 2.8 & 7.5 & 2.8 & 5.6 & 3.2\\ 
& & 20 & 2.9 & 7.7 & 3.0 & 7.2 & 4.4  \\\cline{2-8}
& \multirow{3}{*}{Axial}& 1 & 2.4 & 4.2 & 2.5 & 4.4 & 2.9 \\
& & 10 & 2.5 & 4.4 & 2.5 & 5.2 & 3.1\\ 
& & 20 & 2.3 & 4.1 & 2.7 & 5.1 & 3.3  \\  \hline 
\end{tabularx}
\reducefigurespace
\end{table*}
Table~\ref{tab:SpatialResolution} summarizes the radial, tangential, and axial resolutions, expressed in terms of FWHM and full width tenth maximum (FWTM) for the six source positions at radial distances of \SI{1}{\cm}, \SI{10}{\cm}, and \SI{20}{\cm}, across both configurations with FDR. Additionally, the reported FWHMs for uExplorer are also provided~\cite{spencer2021performance} for reference. For the results without FDR, refer to Fig.~\ref{fig:TPB-Xe-FWHM} and Fig.~\ref{fig:TPB-Xe-FWTM}.
\ThreeDPi\ demonstrates spatial resolutions that are generally comparable to or slightly superior to those of existing PET scanners worldwide, at least in the simulation stage. In the LAr+Xe configuration, the averaged FWHM is \SI{2.7}{\mm} across both axial positions. The FWHMs exhibit minimal variation of \SI{< 1}{mm} across different radial source positions. This consistency indicates that \ThreeDPi\ has the potential to generate precise images regardless of the source's location within the FOV. Despite the initial resolution degradation of the LAr+TPB configuration being double of LAr+Xe without FDR, the subsequent application of FDR yields similar resolutions for both configurations.
The effectiveness of our FDR method is demonstrated through comparisons between raw data (without FDR) and FDR images in Fig.~\ref{fig:RAW-FDR} showcasing resolution enhancement and overall image quality improvement. 
\subsection{Count Rate Performance}
\begin{figure}[h]
\reducefigurespace
\begin{center}
\includegraphics[width=\columnwidth]{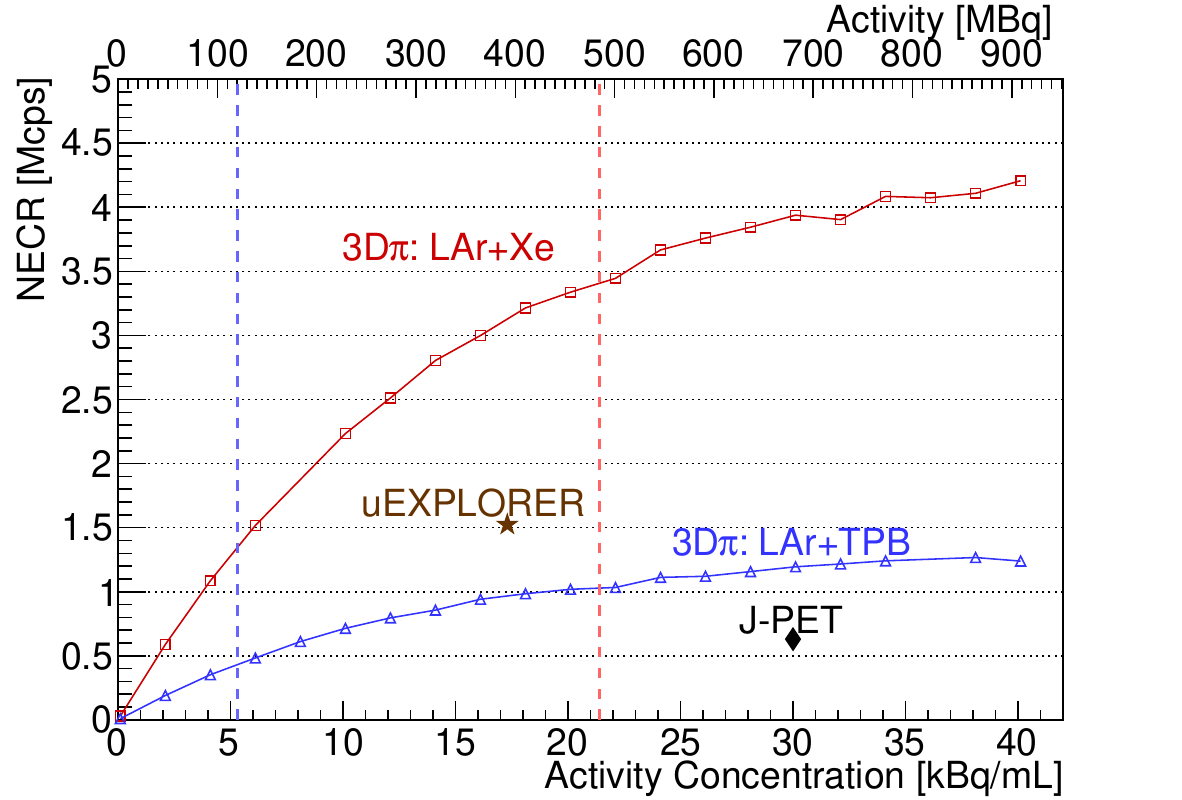}
\end{center}
\reducefigurespace
\caption{Relationship between NECR and activity concentration for both configurations, compared with the peak NECR values of uExplorer~\cite{spencer2021performance} and J-PET~\cite{moskal2021simulating}; {\bf Blue (Red) vertical dashed line:} Activity concentration used as the background (hot spheres) in the Image Quality test.}
\label{fig:NECR}
\reducefigurespace
\reducefigurespace
\end{figure}

The NECR as a function of activity concentration is reported in Fig.~\ref{fig:NECR} for both configurations. The peak NECR values of uEXPLORER~\cite{spencer2021performance} and J-PET~\cite{moskal2021simulating} are included for comparison. The LAr+Xe configuration consistently outperforms the LAr+TPB configuration in NECR values by at least a factor of three at all simulated activity concentrations.

\begin{figure*}[tbh!]
\vspace{-20pt}
\begin{center}
\includegraphics[width=0.99\columnwidth]{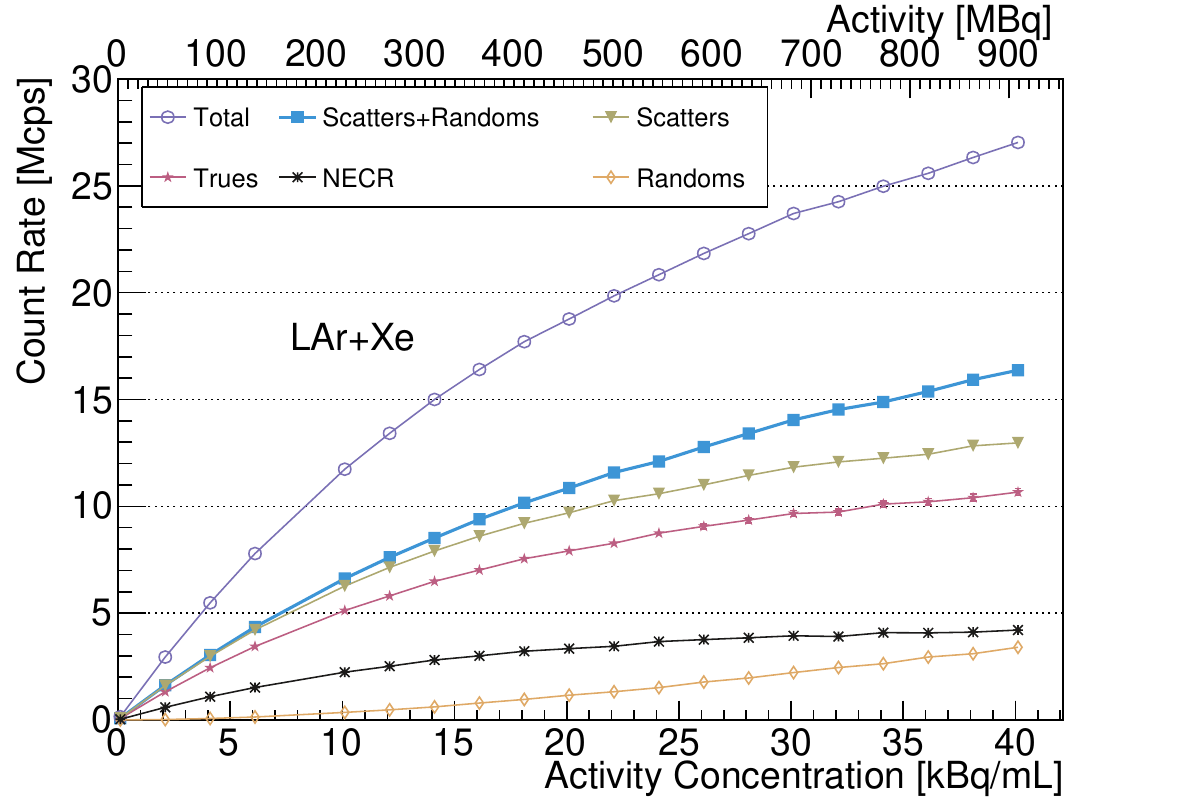}
\includegraphics[width=0.99\columnwidth]{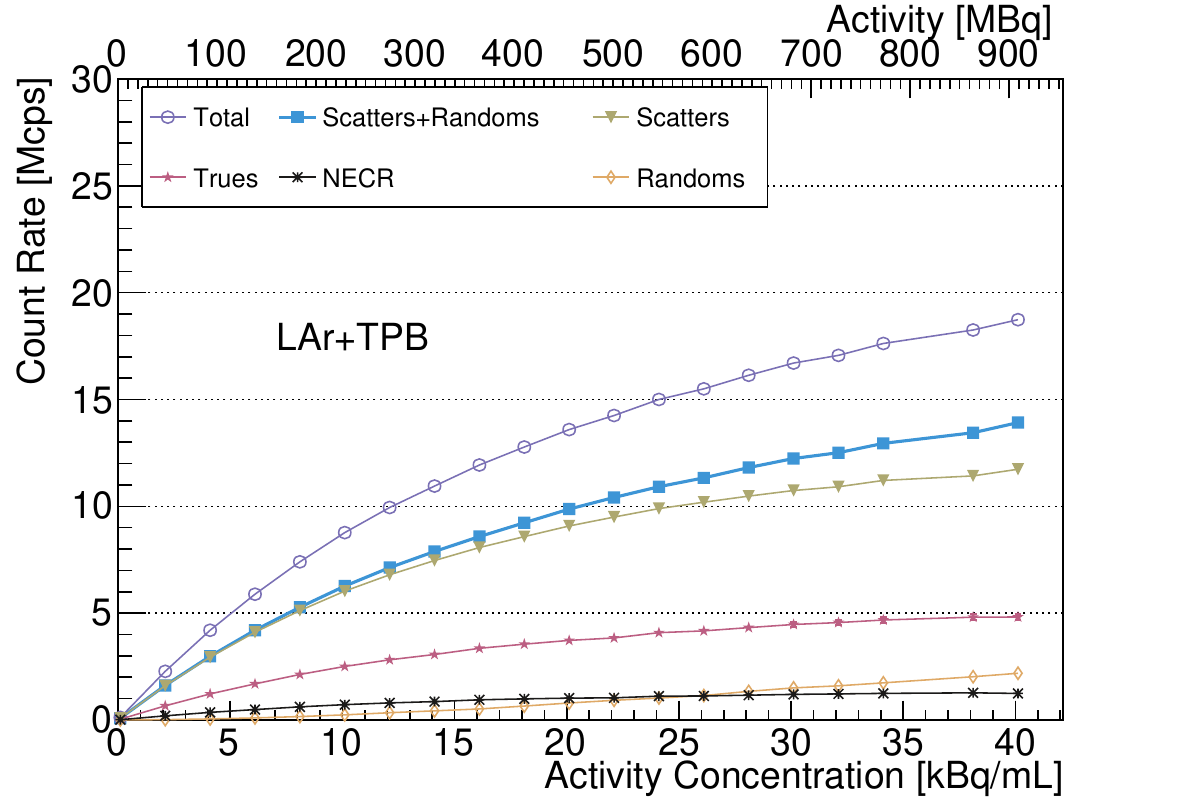}
\end{center}
\reducefigurespaceaftercaption
\caption{Count rates, {\bf Left:} LAr+Xe configuration; {\bf Right:} LAr+TPB configuration.}
\label{fig:countrateXe-TPB}
\vspace{-25pt}
\end{figure*}

As shown in Fig.~\ref{fig:countrateXe-TPB}, the true count rates in the LAr+TPB configuration are almost half of those in LAr+Xe. 
This is because some of the true coincidence events in LAr+Xe are reconstructed far from the line source in the LAr+TPB configuration due to the worse spatial resolution and are categorized as scatter and random events. In the LAr+TPB configuration, the wavelength-shifting process of TPB introduces a delay to the scintillation process. This, coupled with the longer triplet decay component of LAr, significantly degrades the timing resolution and, thus spatial resolution as well. On the other hand, the presence of xenon in the LAr+Xe configuration helps suppress the triplet decay component to around 100 ns. This suppression results in faster scintillation processes, leading to improved timing resolution and higher true count rates than the LAr+TPB configuration. The faster scintillators possess lower dead time, resulting in improved count rate performance, especially at high activity concentrations. This characteristic is particularly evident in the LAr+Xe configuration, where the NECR continues to increase steadily without reaching a peak. In contrast, for the LAr+TPB configuration, a peak NECR of \SI{\sim 1.3}{Mcps} is achieved at \SI{34}{kBq/mL}. Scatter Fraction at similar activity concentrations to uEXPLORER's peak NECR is \SI{51}{\percent} for the LAr+Xe configuration. 
\subsection{OPC Threshold}
The rationale for choosing the OPC threshold is based on evaluating NECR for the LAr+Xe configuration with various OPCs at similar activity concentrations to uEXPLORER's peak NECR (\SI{17.3}{kBq/mL}). With 3000 OPC, NECR reaches a maximum of \SI{3.2}{\Mcps}. It is improved from \SI{2.8}{\Mcps} without OPC. 
As we designed, with the OPC the scatter events are reduced as well as the random events from scatter gammas without losing the true coincident events (see Fig.~\ref{fig:countrateXe-TPB} left side and Fig.~\ref{fig:xe-cut0}).
\subsection{Sensitivity}
\begin{figure}[hbt!]
\begin{center}
\includegraphics[width=0.95\columnwidth]{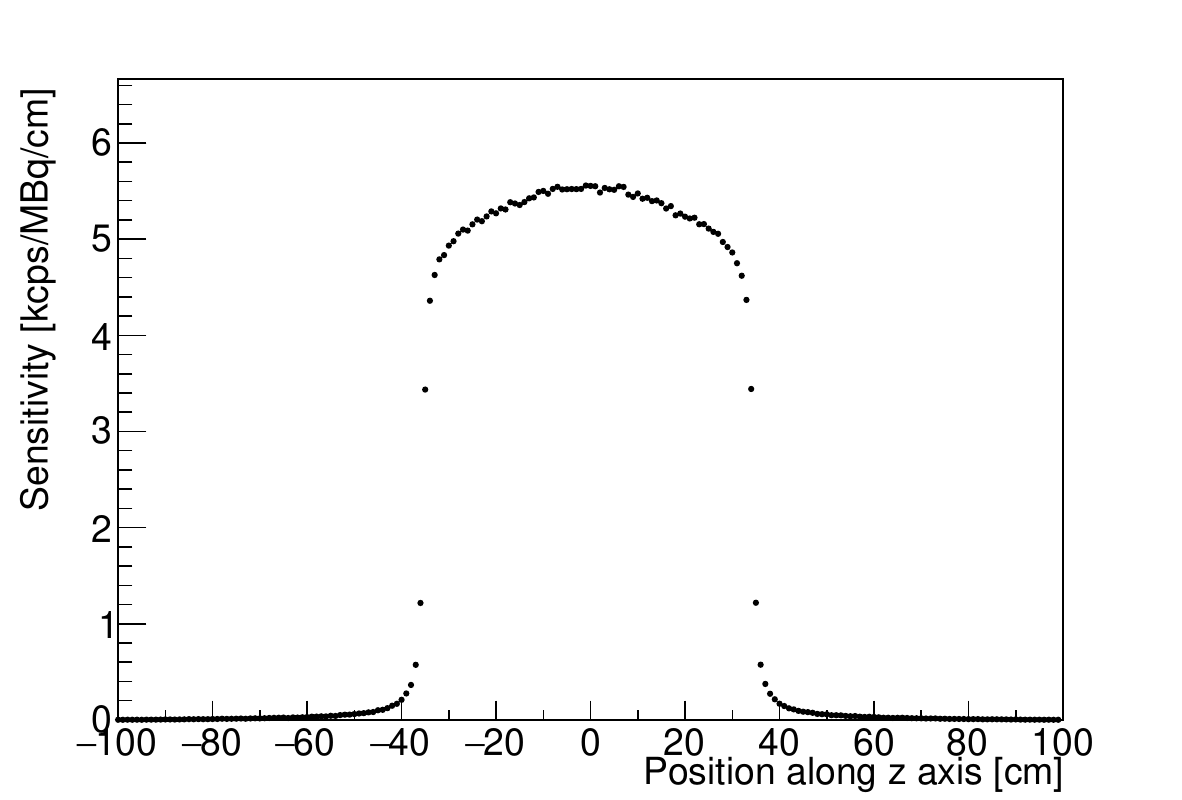}
\end{center}
\reducefigurespaceaftercaption
\caption {Axial sensitivity profile the line source at the center of the scanner; {\bf Configuration:} LAr+Xe.}
\label{fig:Sensy}
\reducefigurespace
\end{figure}

Table~\ref{tab:Syssensitivity} presents a comparison of our system sensitivity in the LAr+Xe configuration with other PET scanners. As expected for TB-PET configurations~\cite{vandenberghe2020state}, our system exhibits significantly higher sensitivity than conventional PET scanners. Although the application of OPC results in a reduction of system sensitivity, our system still outperforms uExplorer~\cite{spencer2021performance} with a sensitivity of more than double.
\begin{table*}[bth!]
\vspace{-10pt}
\centering
\caption{System sensitivity and TOF resolution; {\bf Configuration:} LAr+Xe, *Estimated TOF resolution, see \cite{moskal2021simulating}.}
\label{tab:Syssensitivity}
\begin{tabularx}{1.95\columnwidth}{X X X X X X}
\hline
 & \ThreeDPi & \ThreeDPi (OPC) &  uExplorer~\cite{spencer2021performance}  & J-PET~\cite{moskal2021simulating}  & DMI Gen2 6R~\cite{zeimpekis2022nema} \\
 & \tiny AFOV=200 cm & \tiny  AFOV=200cm & \tiny AFOV=194cm &  \tiny AFOV=200cm &PET/CT\tiny AFOV=30cm \\
\cline{2-6}
Radial position & \multicolumn{5}{c} {System Sensitivity \scriptsize [kcps/MBq]}\\
\Xhline{2pt}
0  cm offset & 571 &373  & 174  & 38 & 32.64  \\ 
10 cm offset & 518 & 347  & 177 &  - & 32.88  \\ \hline
\multicolumn{6}{c} {TOF Resolution \scriptsize [ps]}\\ 
\Xhline{2pt}
At 5.3 kBq/mL &  160  & 151  & 505  & 240*  & 407.6 \\ \hline
\end{tabularx}
\end{table*}
An axial sensitivity profile with a line source at the center of the FOV, is illustrated in Fig.~\ref{fig:Sensy}. The sensitivity profile remains high and constant (\SI{<10}{\percent} variation) across the axial direction.
\subsection{Image Quality}
\begin{figure}[H]
\begin{center}
\includegraphics[width=0.75\columnwidth]{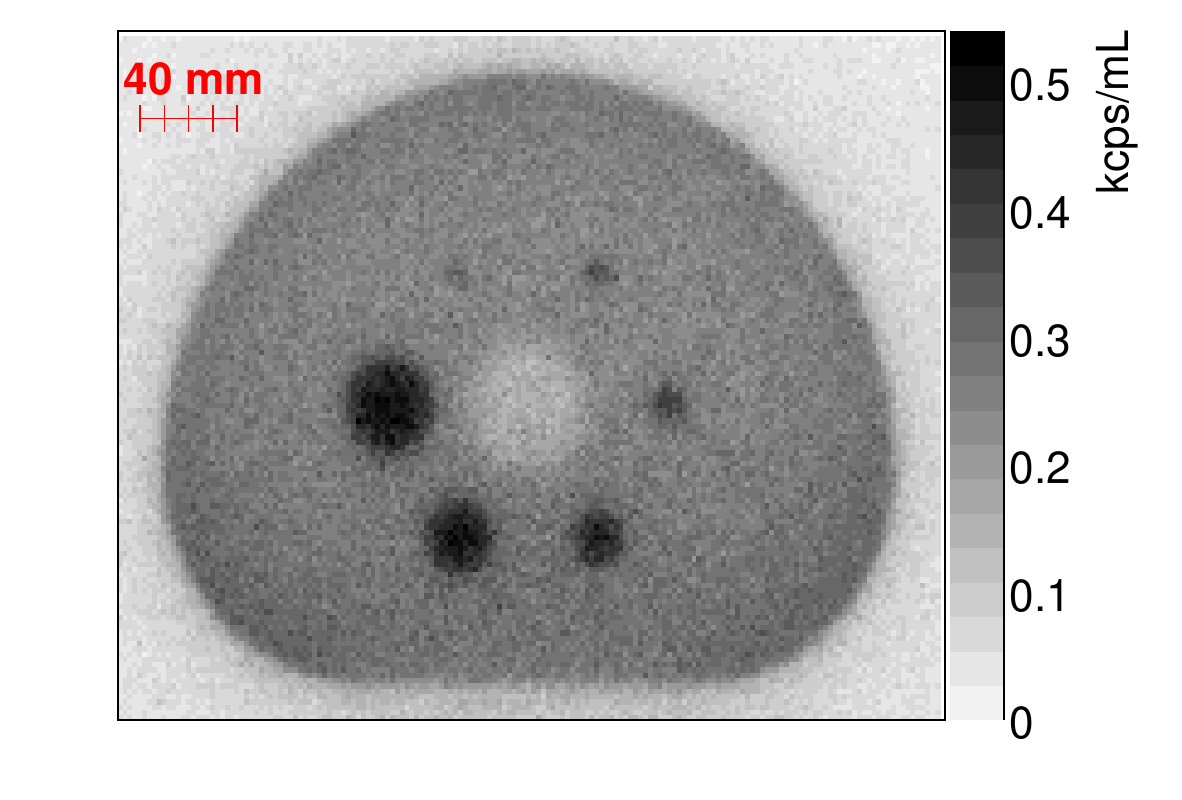}
\vspace{-3.5mm}\includegraphics[width=0.75\columnwidth]{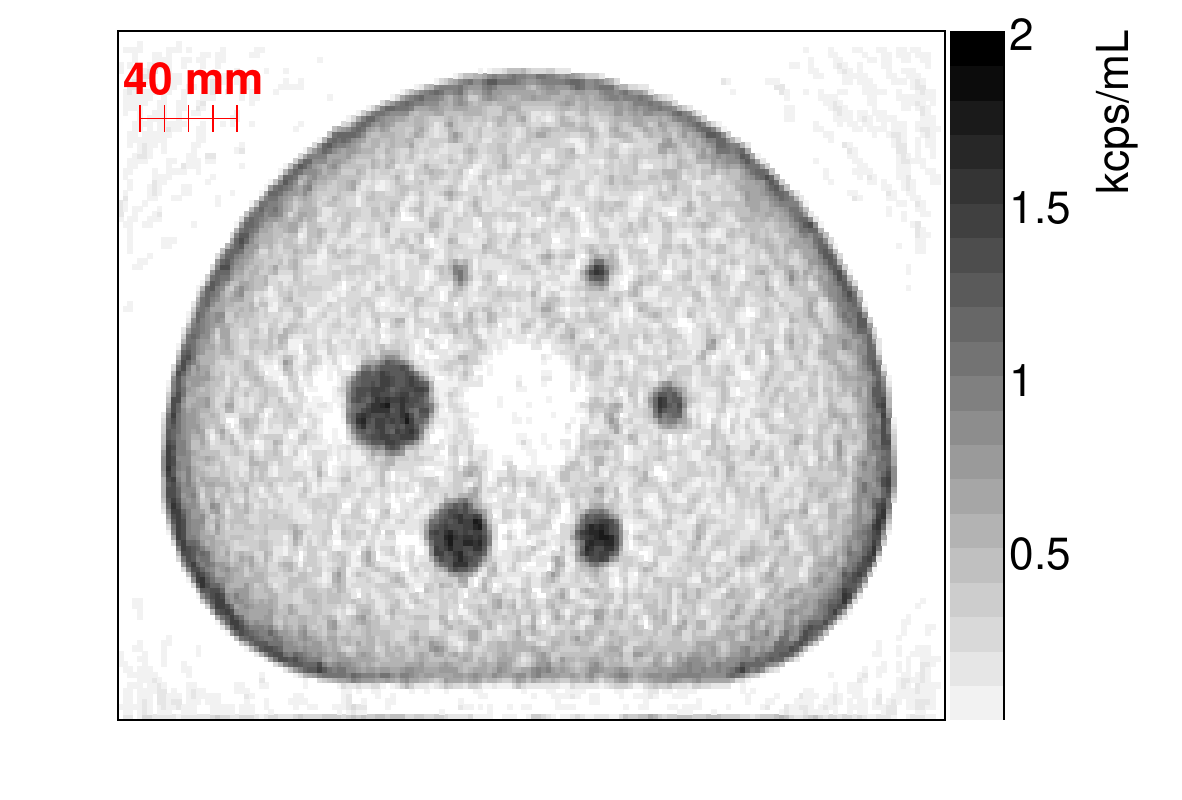}
\end{center}
\vspace{-23pt}
\caption{Central slice of the image quality phantom {\bf Left:} raw data , {\bf Right:} FDR images, {\bf Configuration:} LAr+Xe.}
\label{fig:IQSlice}
\reducefigurespace
\end{figure}

For this test, using the LAr+Xe configuration, we opted for a reduced scanning time of \SI{1}{\minute} instead of \SI{30}{\minute} defined in the NIMA standard. This decision was motivated by the computational demands and storage requirements of generating Monte Carlo simulation data. Due to our high system sensitivity, characterized by an extended AFOV and sufficient NECR, the reduction in scanning time does not compromise the integrity of this test. We assessed the SNR (Eq.~\ref{eq:SNReq},~\cite{lois2010assessment}) as a critical indicator of image clarity to evaluate the performance of \ThreeDPi. 
Compared to the raw data, the FDR method resulted in a more than threefold improvement in SNR (Fig.~\ref{fig:SNR}), demonstrating a significant enhancement in image quality and signal clarity. Figure.~\ref{fig:IQSlice} provides a visual comparison of the 2D central slice of the image quality phantom, further illustrating the improvements achieved through the FDR method.
\begin{table}[bht!]
\reducefigurespace
\centering
\caption{Average relative error, {\bf Configuration:} LAr+Xe.}
\label{tab:IQRE}
\begin{tabularx}{\columnwidth}{X  >{\centering}X  >{\centering}X  >{\centering}X >{\raggedleft\arraybackslash}X}
\hline
 &  \multicolumn{2}{c}{Raw data} &\multicolumn{2}{c}{FDR Image} \\
 \Xhline{2pt}
   & without    &  with   & without   & with  \\
   &  OPC   &  OPC  & OPC  & OPC  \\\hline
RE &  52$\%$ & 39$\%$&  6$\%$&  -24$\%$   \\ \hline 
\end{tabularx}
\reducefigurespace
\end{table}
\begin{table}[thb!]
\reducefigurespace
\centering
\caption{ Percent contrast and background variability; {\bf Configuration:} LAr+Xe.}
\label{tab:IQ}
\begin{tabularx}{\columnwidth}{X >{\centering}X >{\centering}X >{\centering}X >{\raggedleft\arraybackslash}X }
\hline
Sphere  &PC   & PC   & BV  & BV  \\
Diameter & (FDR) & (raw) &(FDR) &(raw) \\
\Xhline{2pt}
10 mm& 68$\%$& 4$\%$& 22$\%$& 5$\%$  \\ 
13 mm& 90$\%$& 8$\%$& 12$\%$& 5$\%$  \\ 
17 mm& 101$\%$& 12$\%$& 8$\%$& 5$\%$ \\ 
22 mm& 126$\%$& 20$\%$& 7$\%$& 5$\%$ \\ 
28 mm& 124$\%$& 23$\%$& 7$\%$& 5$\%$  \\ 
37 mm& 110$\%$& 23$\%$& 7$\%$& 5$\%$  \\ \hline
\end{tabularx}
\reducefigurespace
\end{table}
 Additionally, Table~\ref{tab:IQRE} presents the average RE in the lung insert, where a non-zero value indicates noise or misreconstruction. We observed a reduction in RE after applying OPC, as anticipated, and further improvement with the FDR method. After applying FDR with OPC, we observed negative values for the RE. These negative values indicate a potential over-correction of the background signal during the reconstruction process. Finally, Table~\ref{tab:IQ} presents the PC, which is \SI{100}{\percent} for an ideal image, and percent BV, representing non-uniformity of the image, for both raw data and FDR images, showcasing the effectiveness of our FDR method. The observed PC values exceeding \SI{100}{\percent} can be attributed to the over-correction of the background signal during the reconstruction process. However, the BV increased after applying the FDR method, particularly noticeable for smaller sphere diameters. This increase may be attributed to the algorithm's tendency to over-correct the background signal during reconstruction, yielding a loss of image uniformity and potentially affecting the accuracy of quantitative analyses. 
\subsection{TOF Resolution}
Our study achieved a TOF resolution of \SI{151}{ps} at an activity concentration of \SI{5.3}{kBq/mL} for LAr+Xe configuration. Furthermore, Table~\ref{tab:Syssensitivity}, in its second section, presents a comparison of our TOF resolution with other PET systems, including uExplorer~\cite{spencer2021performance}, J-PET~\cite{moskal2021simulating}, and DMI PET/CT~\cite{zeimpekis2022nema}. TOF resolution improves from \SI{152}{ps} to \SI{136}{ps} as the activity concentration increases from \num{0.1} to \SI{40}{kBq/mL} as shown in Fig.~\ref{fig:TOF}.

\section{Discussion}
\label{sec:Discussion}
While total-body PET imaging is a well-established concept, utilizing a cryogenic scintillator for PET imaging represents a novel and promising approach deserving thorough evaluation. In our study, we focused on assessing the performance of this cryogenic geometry, particularly with the LAr+Xe scintillator, using Monte Carlo simulations to evaluate the performance regarding NEMA NU 2-2018 metrics. Significantly, our analysis underscores the consistent superiority of the LAr+Xe configuration over the LAr+TPB counterpart. The count rate values as depicted in both panels of Fig.~\ref{fig:countrateXe-TPB} and the spatial resolution based on raw data in Fig.~\ref{fig:TPB-Xe-FWHM} and Fig.~\ref{fig:TPB-Xe-FWTM} reveal a remarkable enhancement in system performance. The LAr+Xe configuration exhibits more than a two-fold improvement in spatial resolution and a three-fold increase in NECR (Fig.~\ref{fig:NECR}) at all simulated activity concentrations. This superior performance can be attributed to the pivotal role of xenon doping, which effectively suppresses the triplet decay component and facilitates faster scintillation processes. The inherent delay introduced by TPB in the scintillation process of LAr+TPB degrades the overall timing resolution and compromises the system's ability to capture rapid events accurately. As a caveat, the geometry of this simulation as described in section~\hyperref[ssec:Geometry]{\ThreeDPi\ Geometry} is rather simplistic compared to lab-based actual scanners. Thus, we expect the number of scatters in the geometry to increase with additional layers of material subsequently lowering our NECR. These results demonstrate the efficacy of the chosen configuration in optimizing system performance, reaffirming the importance of xenon doping in cryogenic scintillator-based PET imaging. The ability of the LAr+Xe configuration to maintain high NECR values across a range of activity concentrations highlights its suitability for clinical applications where high count rates and superior timing resolution are essential for accurate imaging. 

However, despite its promising advantages, the LAr+Xe configuration poses its own set of challenges, particularly in achieving long-term operational stability and homogeneity of the LAr+Xe mixture. Addressing these challenges through future research and development will be crucial to fully realize the potential of xenon doping in cryogenic scintillator-based PET imaging systems. Furthermore, the sensitivity of a PET scanner is profoundly influenced by its geometry, particularly the AFOV. Our comparison in Tables~\ref{tab:Syssensitivity} highlights the superior sensitivity of \ThreeDPi\ compared to scanners with shorter or similar AFOV lengths. This enhancement owes itself to the innovative multiple-layer structure, incorporating concentric layers of LAr+Xe scintillators. This design effectively increases the amount of gamma rays collected by the PET scanner, substantially boosting sensitivity. 

The unique characteristics of \ThreeDPi, particularly its monolithic and fast scintillator, present opportunities for exploring novel reconstruction algorithms. Each detection layer in \ThreeDPi\ utilizes outer and inner layers of SiPMs to mitigate depth of interaction (DOI) parallax error, a common issue in PET scanners leading to spatial measurement uncertainties. While various scintillator configurations exist for DOI measurement, the outer/inner SiPM setup was chosen for its ease of implementation in simulations for cryogenic scintillators. Moreover, this configuration enables precise localization of each gamma interaction without requiring a segmented scintillator. While the multi-layer design of \ThreeDPi\ offers advantages in sensitivity and spatial resolution, it also introduces complexity and cost. Especially, the proper front-end electronics need to be developed to handle a large number of channels for the fast scintillation signal at cryogenic temperatures. Future studies will focus on optimizing the geometry to balance performance with feasibility and cost-effectiveness for practical implementation. Consequently, the optimization of the scanner's geometry will be a primary focus of upcoming research projects, particularly in the development of a prototype.

Furthermore, the FDR employed \SigmaPar\ and \SigmaPerp\ for the resolution of LOR, and TOF information, facilitating precise localization of annihilation points. Despite not representing the most advanced image reconstruction technique and using a reduced scanning time, FDR was also successfully implemented to showcase the potential of \ThreeDPi\ in achieving clinically relevant image quality. FDR was successful in significantly improving SNR by more than three-fold compared to raw data (Fig.~\ref{fig:SNR}), highlighting its utility in \ThreeDPi\ image reconstruction. The decrease in RE confirms its effectiveness in reducing noise and improving image reconstruction accuracy. However, the negative RE values observed after applying FDR with an OPC suggest a potential over-correction of the background signal, which warrants further investigation.

Moving forward, future research developments will also focus on enhancing our reconstruction algorithms to address challenges like the potential over-correction of background signals. Additionally, efforts will be directed towards improving the overall performance and stability of the \ThreeDPi\ system, including optimizing the geometry and exploring advanced scintillator configurations. The development of a prototype will be a critical step in understanding and addressing the challenges associated with the design and implementation of \ThreeDPi\ in real-world scenarios.

\section{Conclusions}
\label{sec:Conclusions}
The development and evaluation of the \ThreeDPi\ methodology, validated through Monte Carlo simulations following NEMA NU 2-2018 protocols, represent a significant advancement in PET imaging.  By combining and integrating a homogenous, monolithic scintillator composed of LAr with doped Xe, \ThreeDPi\ shows exceptional sensitivity (\SI{373}{kcps/MBq}), superior spatial resolution of \SI{2.7}{\mm} FWHM, and a good TOF resolution of \SI{151}{\ps} at \SI{5.3}{kBq/mL}. Notably, the NECR continued to increase steadily without reaching a peak up to \SI{4.1}{\Mcps} with activity concentrations of \SI{40}{kBq/mL}, indicating \ThreeDPi's ability to handle higher activity concentrations. These findings suggest the potential for shorter scan times and reduced radiation exposure for patients. Xenon doping plays a crucial role by enabling faster scintillation processes, leading to improved timing resolution and reduced noise. While further research is needed to optimize the system's geometry, refine reconstruction algorithms, and address challenges for clinical translation, \ThreeDPi\ holds promise for surpassing existing PET technology and enhancing diagnostic accuracy with improved patient comfort.
\section{Acknowledgements}
\label{sec:Acknowledgements}
Funding for this work was provided by multiple sources, including the International Research Agenda Programme AstroCeNT (GrantNo.~MAB/2018/7) funded from the Foundation for Polish Science from the European Regional Development Fund, the European Union’s Horizon 2020 research and innovation program (952480 DarkWave), the National Science Foundation Graduate Research Fellowship (award 2136513), the University of Houston's Houston Scholars society, INFN, Università degli Studi di Cagliari, Fondazione CON IL SUD, and the resources of the uHPC cluster (acquired through NSF Award Number 1531814). No potential conflicts of interest relevant to this article exist.

\bibliographystyle{./ama}
\bibliography{./ds}

\begin{thebibliography}{10}

\bibitem{Agnes:2016fz}
Agnes P, Agostino L, Albuquerque IFM, et al. {Results from the first use of low radioactivity argon in a dark matter search}  {\it Phys Rev D. } 2016;93(8):081101.

\bibitem{8113575}
D’Incecco M, Galbiati C, Giovanetti GK, et al. Development of a Novel Single-Channel, 24 cm2, SiPM-Based, Cryogenic Photodetector  {\it IEEE Trans Nucl Sci. } 2018;65:591-596.

\bibitem{Aalseth:2017hu}
Aalseth CE, Acerbi F, Agnes P, et al. {Cryogenic Characterization of FBK RGB-HD SiPMs}  {\it J Instrum. } 2017;12(09):9030.

\bibitem{Kubota:1978bi}
Kubota S, Hishida M, Nohara A. Variation of scintillation decay in liquid argon excited by electrons and alpha particles  {\it Nucl Instrume Methods. } 1978;150(3):561-564.

\bibitem{Mao_2013}
Mao R, Wu~C, Dai LE, Lu~S. Crystal growth and scintillation properties of LSO and LYSO crystals  {\it J Cryst Growth. } 2013;368:97-100.

\bibitem{DOKE1990617}
Doke T, Masuda K, Shibamura E. Estimation of absolute photon yields in liquid argon and xenon for relativistic (1 MeV) electrons  {\it Nucl Instrum Methods Phys Res A. } 1990;291(3):617-620.

\bibitem{Pershing_2022}
Pershing T, Xu~J, Bernard E, et al. Performance of Hamamatsu VUV4 SiPMs for detecting liquid argon scintillation  {\it J Instrum.. } 2022;17(04):4017.

\bibitem{Flournoy:1994bx}
Flournoy JM, Berlman IB, Rickborn B, Harrison R. {Substituted tetraphenylbutadienes as fast scintillator solutes}  {\it Nucl Instrum Methods Phys Res A. } 1994;351(2-3):349-358.

\bibitem{KUBOTA199371}
Kubota S, Hishida M, Himi S, Suzuki J, Ruan J. The suppression of the slow component in xenon-doped liquid argon scintillation  {\it Nucl Instrum Methods Phys Res A. } 1993;327(1):71-74.

\bibitem{Cheshnovsky1972}
Cheshnovsky O, Raz B, Jortner J. {Emission Spectra of Deep Impurity States in Solid and Liquid Rare Gas Alloys}  {\it J Chem Phys. } 1972;57(11):4628-4632.

\bibitem{Ootani:2015ga}
Ootani W, Ieki K, Iwamoto T, et al. {Development of deep-UV sensitive MPPC for liquid xenon scintillation detector}  {\it Nucl Instrum Methods Phys Res A. } 2015;787:220-223.

\bibitem{lavoie1976liquid}
Lavoie L. Liquid xenon scintillators for imaging of positron emitters  {\it Med Phys. } 1976;3(5):283-293.

\bibitem{ROMOLUQUE2020162397}
Romo-Luque C. PETALO: Time-of-Flight PET with liquid xenon  {\it Nucl Instrum Methods Phys Res A. } 2020;958:162397.
\newblock Proceedings of the Vienna Conference on Instrumentation 2019.

\bibitem{Galbiati_2021}
Galbiati C, Li~X, Luo J, Marlow DR, Wang H, Wang Y. Pulse shape study of the fast scintillation light emitted from xenon-doped liquid argon using silicon photomultipliers  {\it J Instrum.. } 2021;16(02):02015.

\bibitem{PhysRevC108045503}
Bernard EP, Mizrachi E, Kingston J, et al. Thermodynamic stability of xenon-doped liquid argon detectors  {\it Phys Rev C. } 2023;108(4):045503.

\bibitem{Gundacker_2020}
Gundacker S, Heering A. The silicon photomultiplier: fundamentals and applications of a modern solid-state photon detector  {\it Phys Med Biol. } 2020;65(17):17TR01.

\bibitem{Acerbi:2015ca}
Acerbi F, Ferri A, Zappala G, et al. {NUV Silicon Photomultipliers With High Detection Efficiency and Reduced Delayed Correlated-Noise}  {\it IEEE Trans Nucl Sci. } 2015;62(3):1318-1325.

\bibitem{Acerbi:2017gy}
Acerbi F, Davini S, Ferri A, et al. {Cryogenic Characterization of FBK HD Near-UV Sensitive SiPMs}  {\it IEEE Trans Elec Dev. } 2017;64(2):521-526.

\bibitem{Fossum:1997jp}
Fossum ER. {CMOS image sensors: electronic camera-on-a-chip}  {\it IEEE Trans. Elec. Dev. } 1997;44(10):1689-1698.

\bibitem{Agnes:2017cz}
Agnes P, Albuquerque IFM, Alexander T, et al. {Simulation of argon response and light detection in the DarkSide-50 dual phase TPC}  {\it J Instrum. } 2017;12(10):10015.

\bibitem{Agostinelli:2003fg}
Agostinelli S, Allison J, Amako K, et al. {Geant4-a simulation toolkit}  {\it Nucl Instrum Methods Phys Res A. } 2003;506(3):250-303.

\bibitem{gola2019nuv}
Gola A, Acerbi F, Capasso M, et al. {NUV}-{Sensitive} {Silicon} {Photomultiplier} {Technologies} {Developed} at {Fondazione} {Bruno} {Kessler}  {\it Sensors. } 2019;19(2):308.

\bibitem{national2018nu}
NEMA Standards Publication NU 2-2018: Performance Measurements of Positron Emission Tomographs (PET);   2018.

\bibitem{spencer2021performance}
Spencer B, Berg E, Schmall J, et al. Performance evaluation of the {uEXPLORER} {Total}-body {PET}/{CT} scanner based on {NEMA} {NU} 2-2018 with additional tests to characterize long axial field-of-view {PET} scanners  {\it J Nucl Med. } 2020;62(6):861-870.

\bibitem{moskal2021simulating}
Moskal P, Shopa RY, Raczyński L, et al. Simulating {NEMA} characteristics of the modular total-body {J}-{PET} scanner—an economic total-body {PET} from plastic scintillators  {\it Phys Med Biol. } 2021;66(17).

\bibitem{vandenberghe2020state}
Vandenberghe S, Moskal P, Karp JS. State of the art in total body {PET}  {\it EJNMMI Phys. } 2020;7(1):35.

\bibitem{zeimpekis2022nema}
Zeimpekis KG, Kotasidis FA, Huellner M, Nemirovsky A, Kaufmann PA, Treyer V. {NEMA} {NU} 2–2018 performance evaluation of a new generation 30-cm axial field-of-view {Discovery} {MI} {PET}/{CT}  {\it Eur J Nucl Med Mol Imaging. } 2022;49(9):3023-3032.

\bibitem{lois2010assessment}
Surti S, Karp JS, Kinahan PE, et al. An Assessment of the Impact of Incorporating Time-of-Flight Information Into Clinical PET/CT Imaging  {\it J Nucl Med. } 2010;51(2):237-245.

\end{thebibliography}
\clearpage 
\clearpage 

\fontsize{10pt}{10pt}\selectfont
\renewcommand{\thefigure}{S\arabic{figure}}
\renewcommand{\thetable}{S\arabic{table}}
\renewcommand{\theequation}{S\arabic{equation}} 
\setcounter{page}{1}
\renewcommand{\thepage}{S\arabic{page}}
\section{Supplemental}
\label{sec:Supp}
\begin{figure}[H]
\begin{center}
\includegraphics[width=\columnwidth]{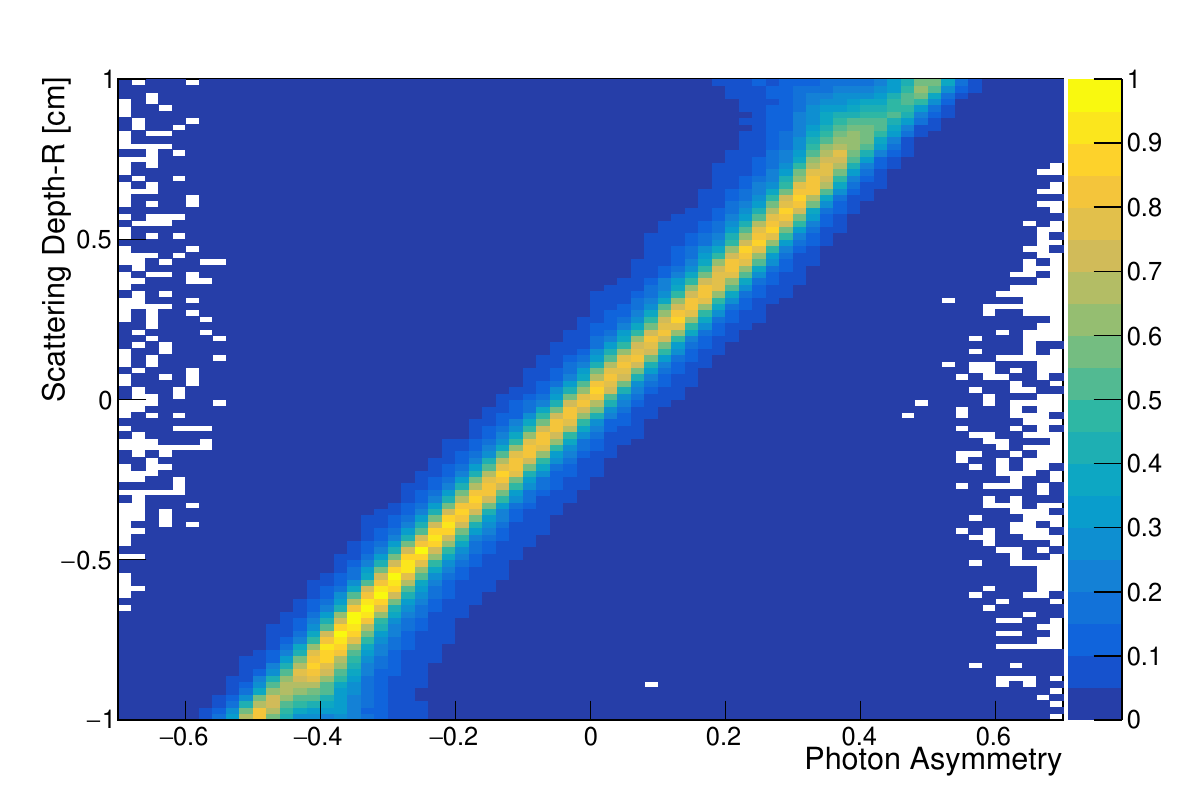}
\caption{Relationship between photon asymmetry and scattering depth: The vertical axis represents the position of the scattering depth of the \grs\, where the inner side corresponds to \SI{-1}{\cm}, the center corresponds to \SI{0}{\cm}, and the outer side corresponds to \SI{1}{\cm}.  The horizontal axis shows the asymmetry of the number of photons detected by the two $3\times3$ SiPMs matrices as defined in the text. Results from LAr+Xe and LAr+TPB configurations are similar.}
\label{fig:Asymmetry}
\end{center}
\end{figure}

\begin{figure}[H]
\begin{center}
\includegraphics[width=\columnwidth]{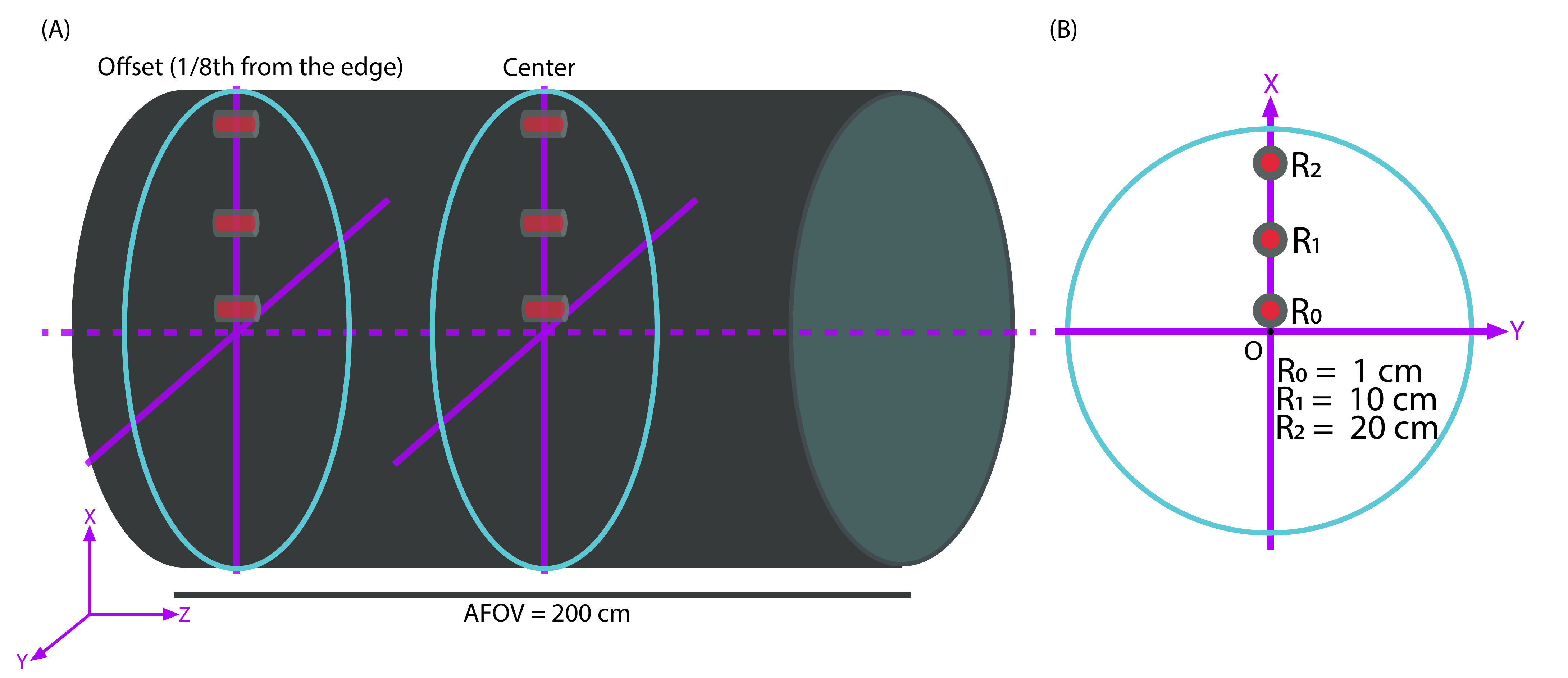}
\end{center}
\caption{Radial distribution of point sources for both axial positions in the NEMA NU 2-2018 spatial resolution test.}
\label{fig:NEMA-Spatial-Res-Geometry}
\end{figure}

\begin{figure}[H]
\begin{center}
\includegraphics[width=\columnwidth]{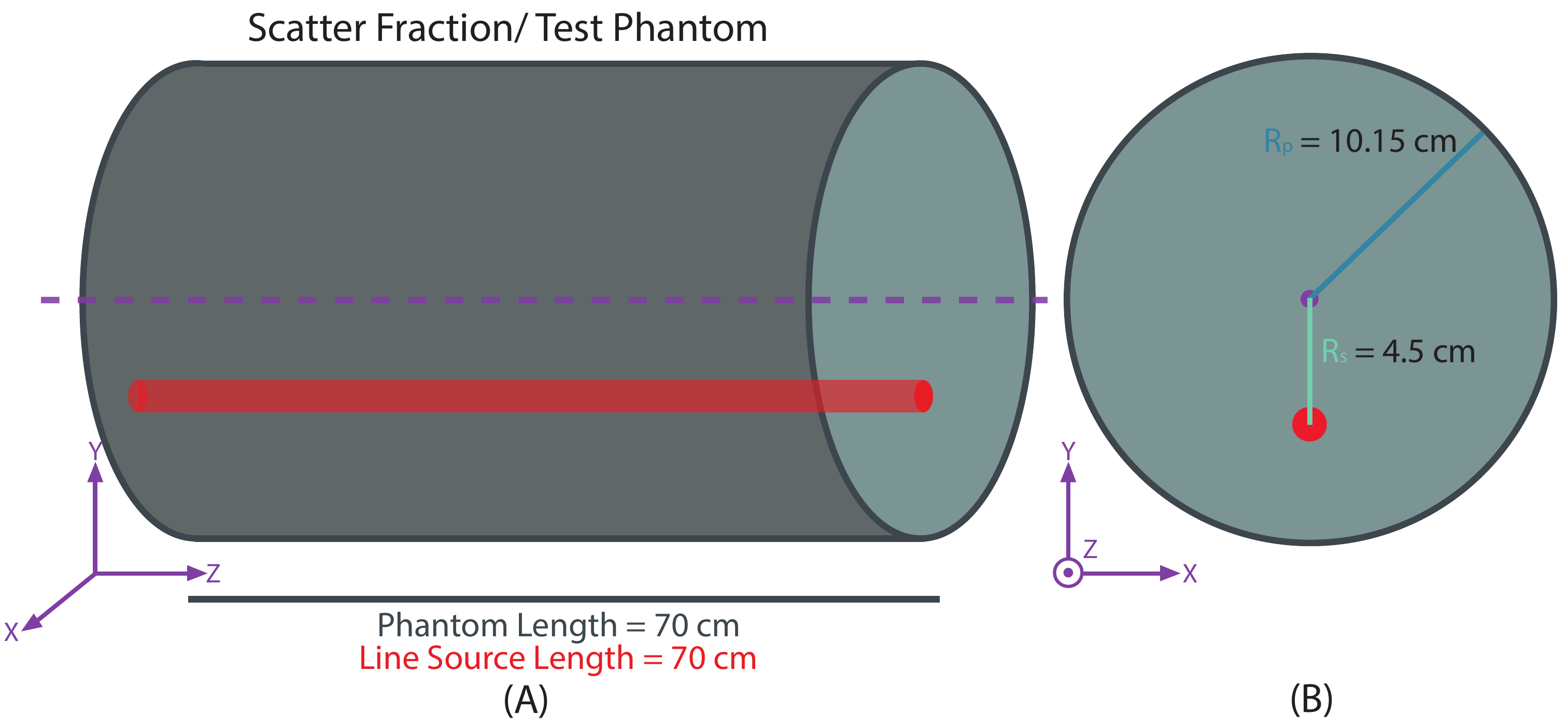}
\end{center}
\caption{Test Phantom and line source for count rate performance based on NEMA NU 2-2018.}
\label{fig:SFPh}
\end{figure}

\begin{figure}[H]
\begin{center}
\includegraphics[width=\columnwidth]{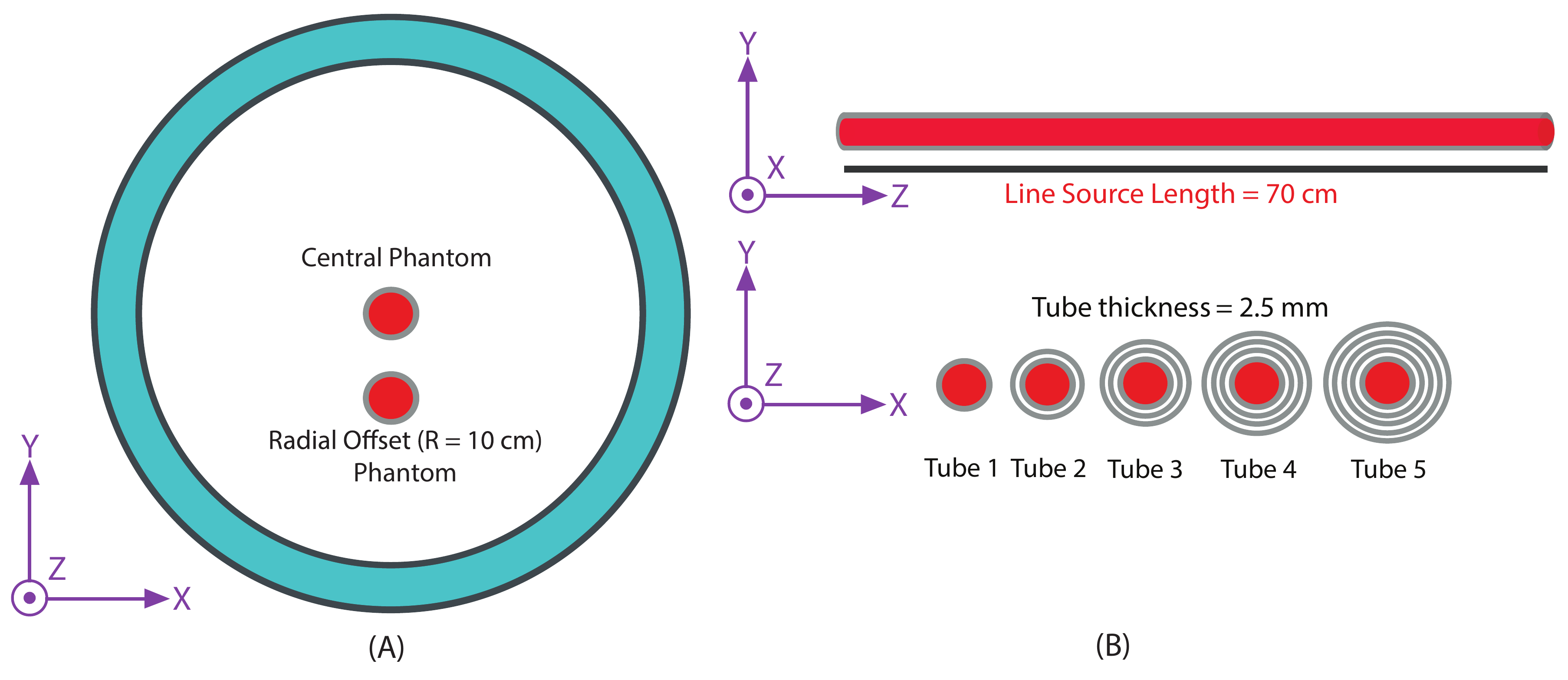}
\end{center}
\caption{Sensitivity Phantoms and line source and their radial positions based on NEMA NU 2-2018.}
\label{fig:SPh}
\end{figure} 

\begin{figure}[H]
\begin{center}
\includegraphics[width=\columnwidth]{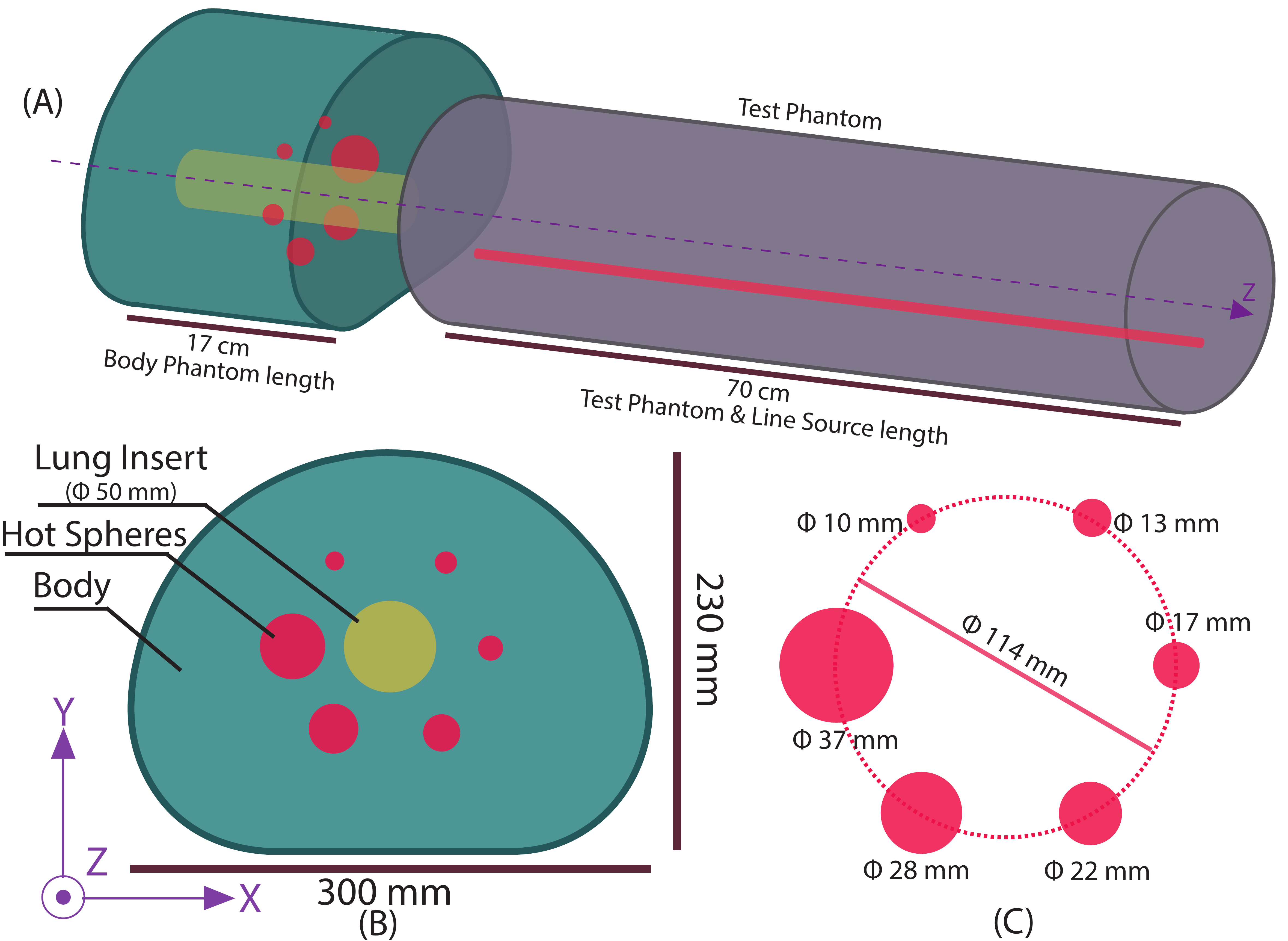}
\end{center}
\caption{Image Quality Phantoms based on NEMA NU 2-2018.}
\label{fig:IQPh}
\end{figure} 

\begin{figure*}[hbt!]
\begin{center}
\includegraphics[width=\columnwidth]{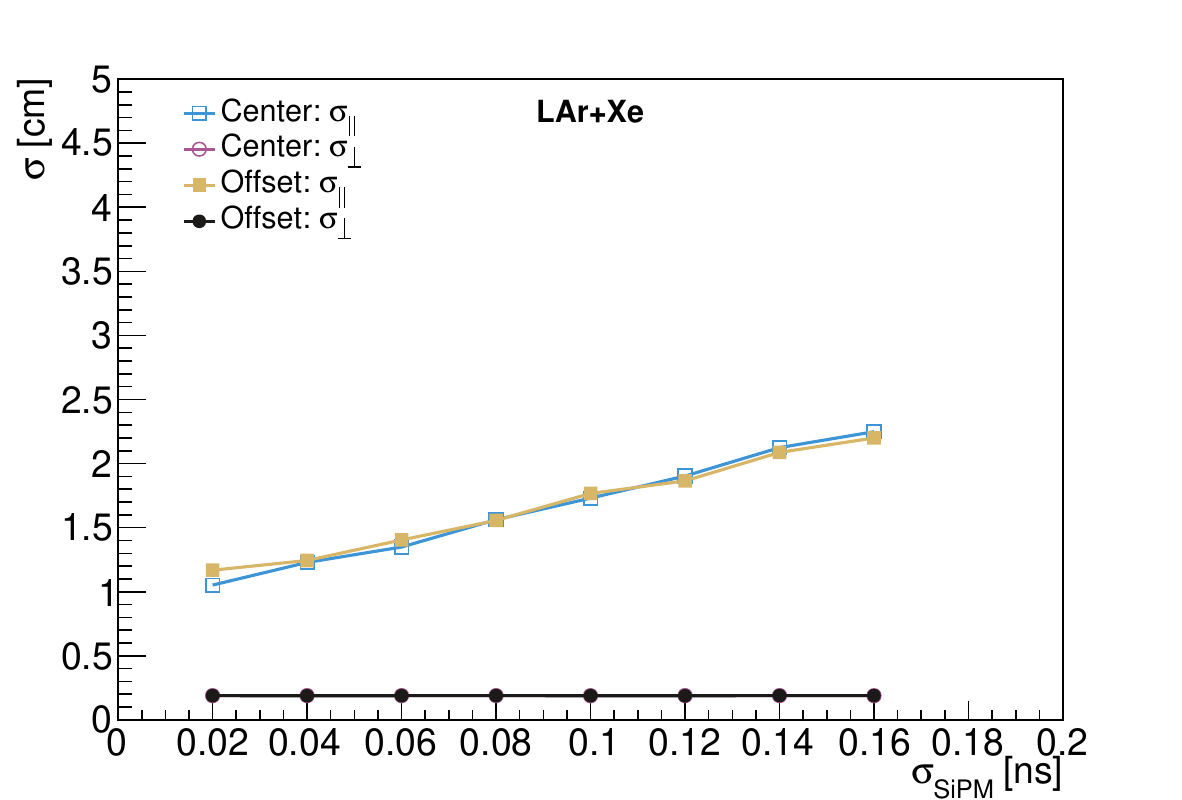}
\includegraphics[width=\columnwidth]{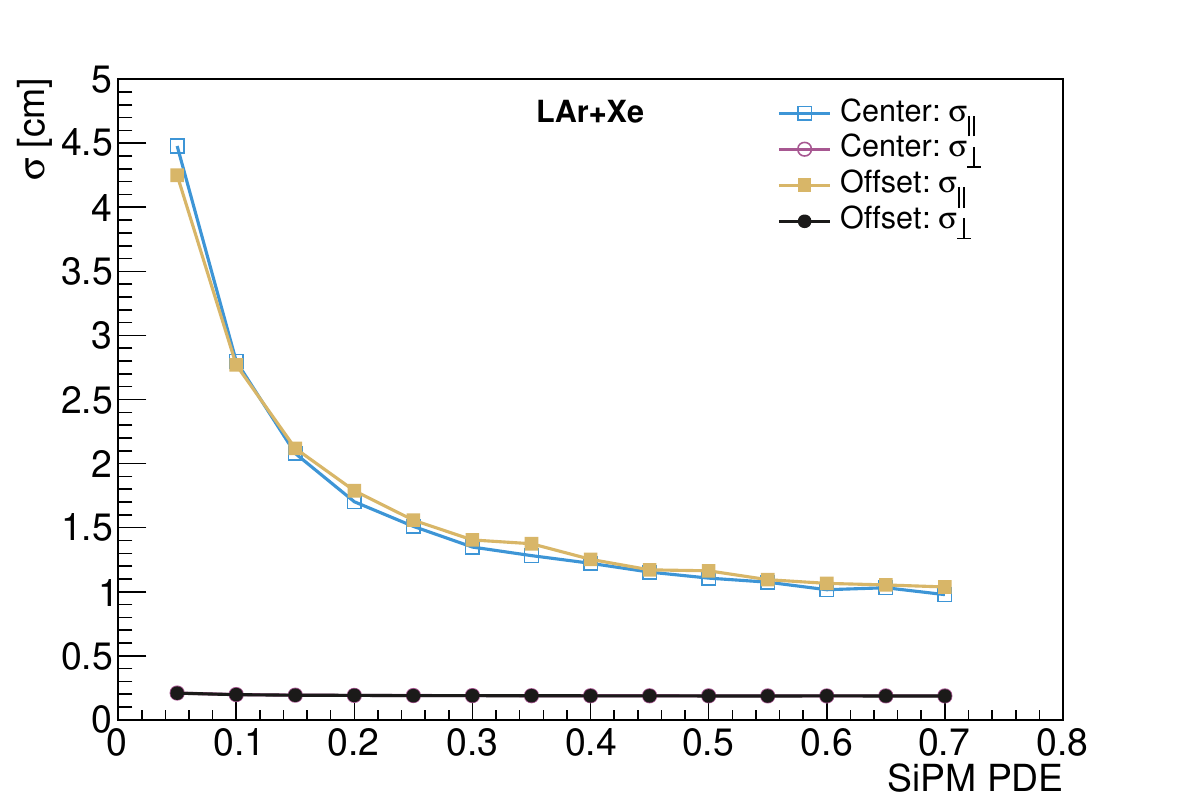}
\end{center}
\caption{{\bf Configuration:} LAr+Xe and without OPC and FDR; source radial position at $r = \SI{1}{\cm}$, Center: center of the AFOV, offset: one-eighth of the AFOV from the end of the tomography device, along the axial axil; Spatial resolution, in directions parallel and perpendicular to LOR, {\bf Left:} vs. \SigmaSiPM.  Parameter: $PDE = 0.30$.  {\bf Right:} vs. SiPM PDE. Parameter: \SigmaSiPM=\SI{60}{\ps}.}
\label{fig:Xe-SigmasVsSiPM}
\end{figure*}

\begin{figure*}[hbt!]
\begin{center}
\includegraphics[width=\columnwidth]{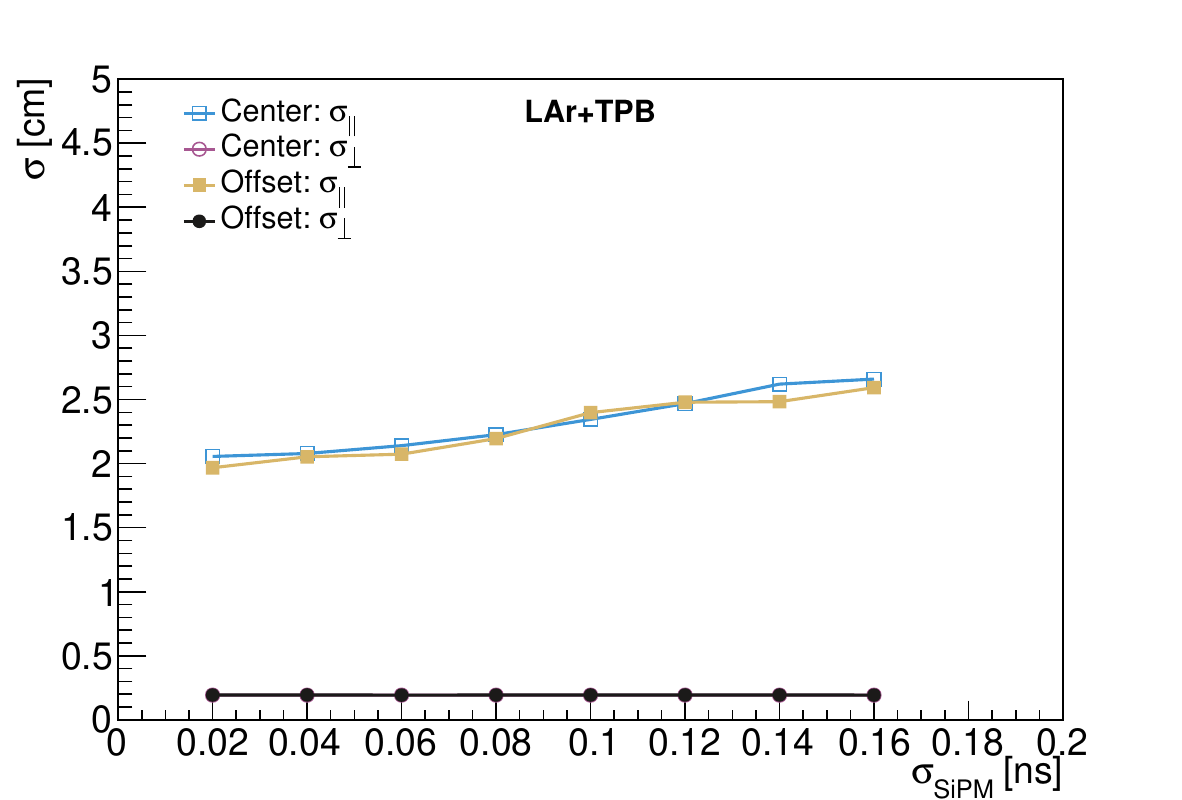}
\includegraphics[width=\columnwidth]{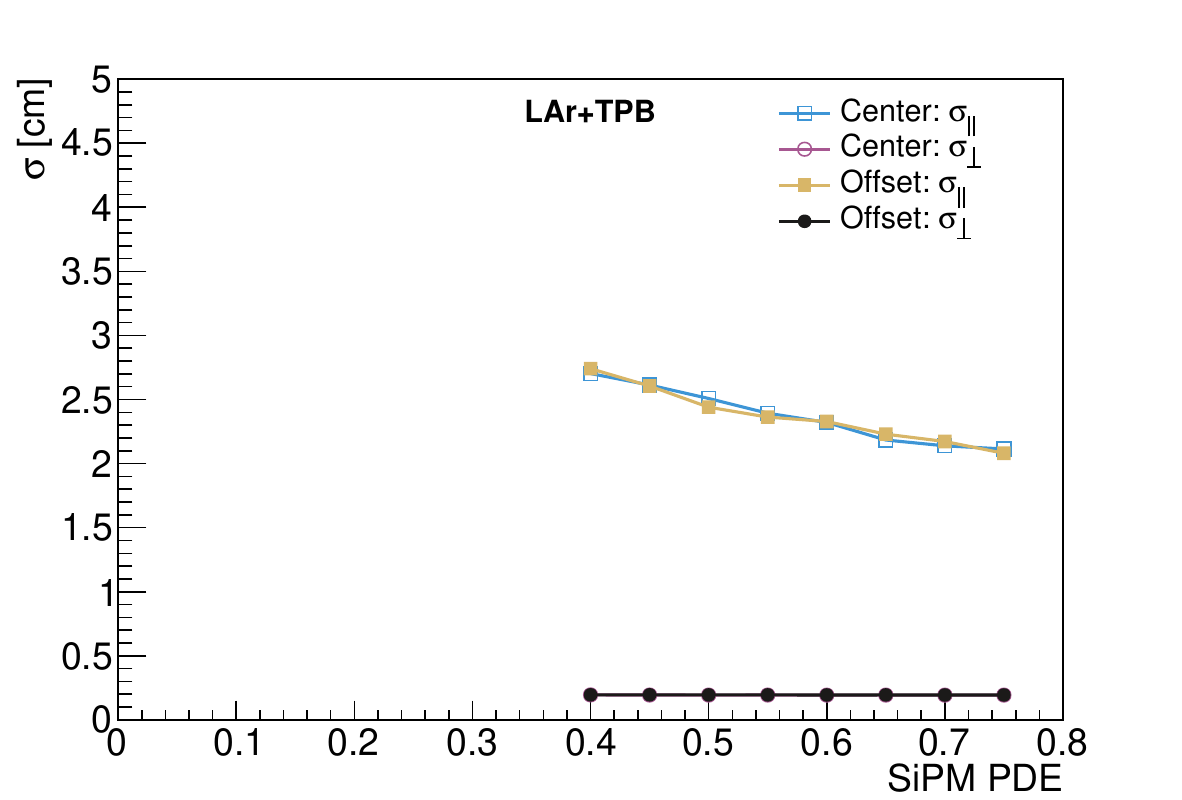}
\end{center}
\caption{{\bf Configuration:} LAr+TPB and without OPC and FDR; source radial position at $r = \SI{1}{\cm}$, Center: center of the AFOV, offset: one-eighth of the AFOV from the end of the tomography device, along the axial axil; Spatial resolution, in directions parallel and perpendicular to LOR, {\bf Left:} vs. \SigmaSiPM.  Parameter: $PDE = 0.70$.  {\bf Right:} vs. SiPM PDE. Parameter: \SigmaSiPM=\SI{60}{\ps}.}
\label{fig:TPB-SigmasVsSiPM}
\end{figure*}

\begin{figure}[H]
\begin{center}
\includegraphics[width=0.9\columnwidth]{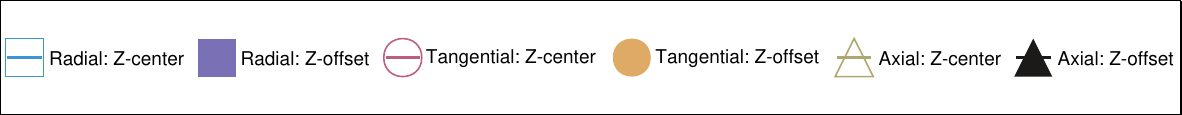}
\includegraphics[width=0.45\columnwidth]{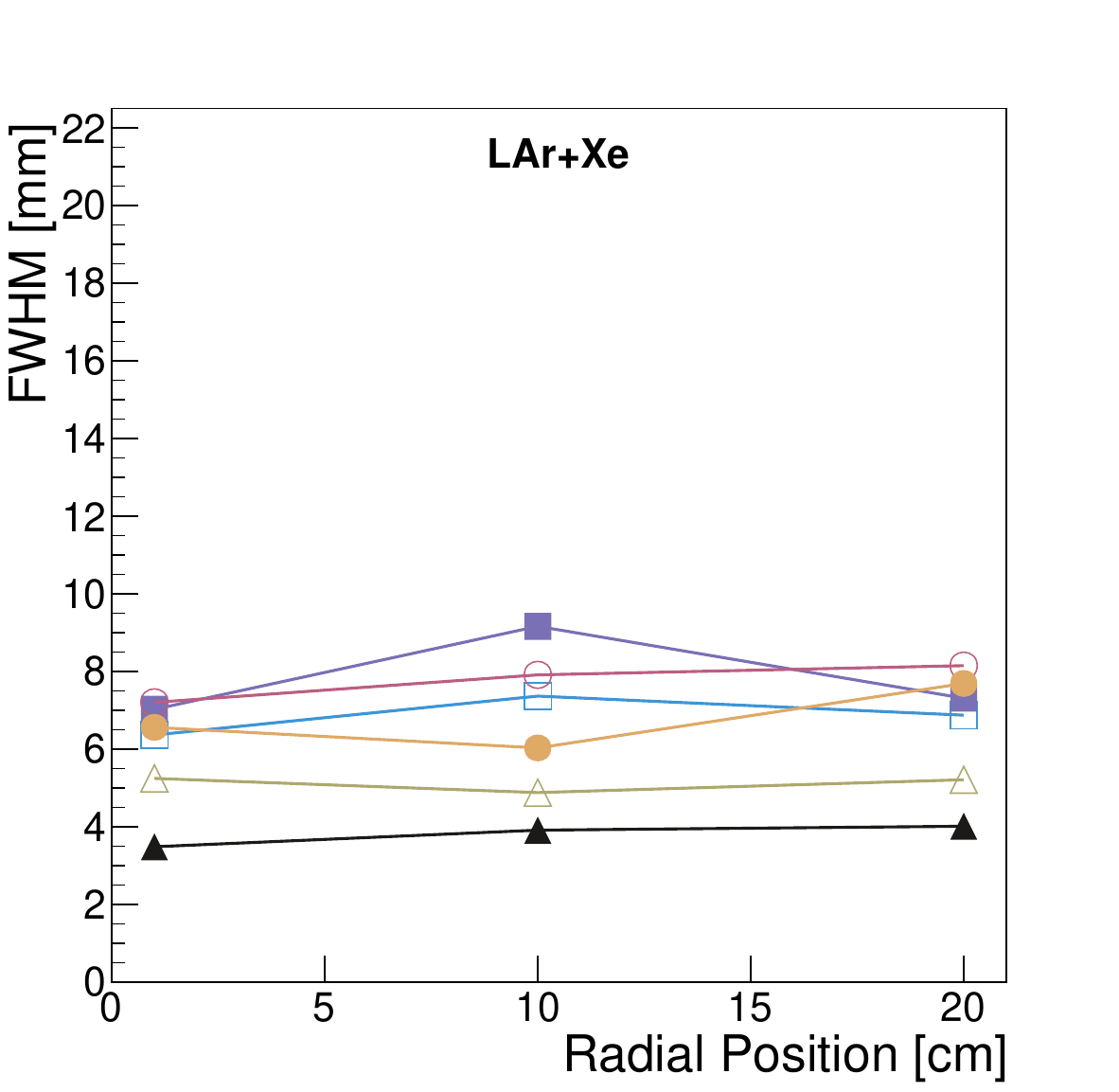}
\includegraphics[width=0.45\columnwidth]{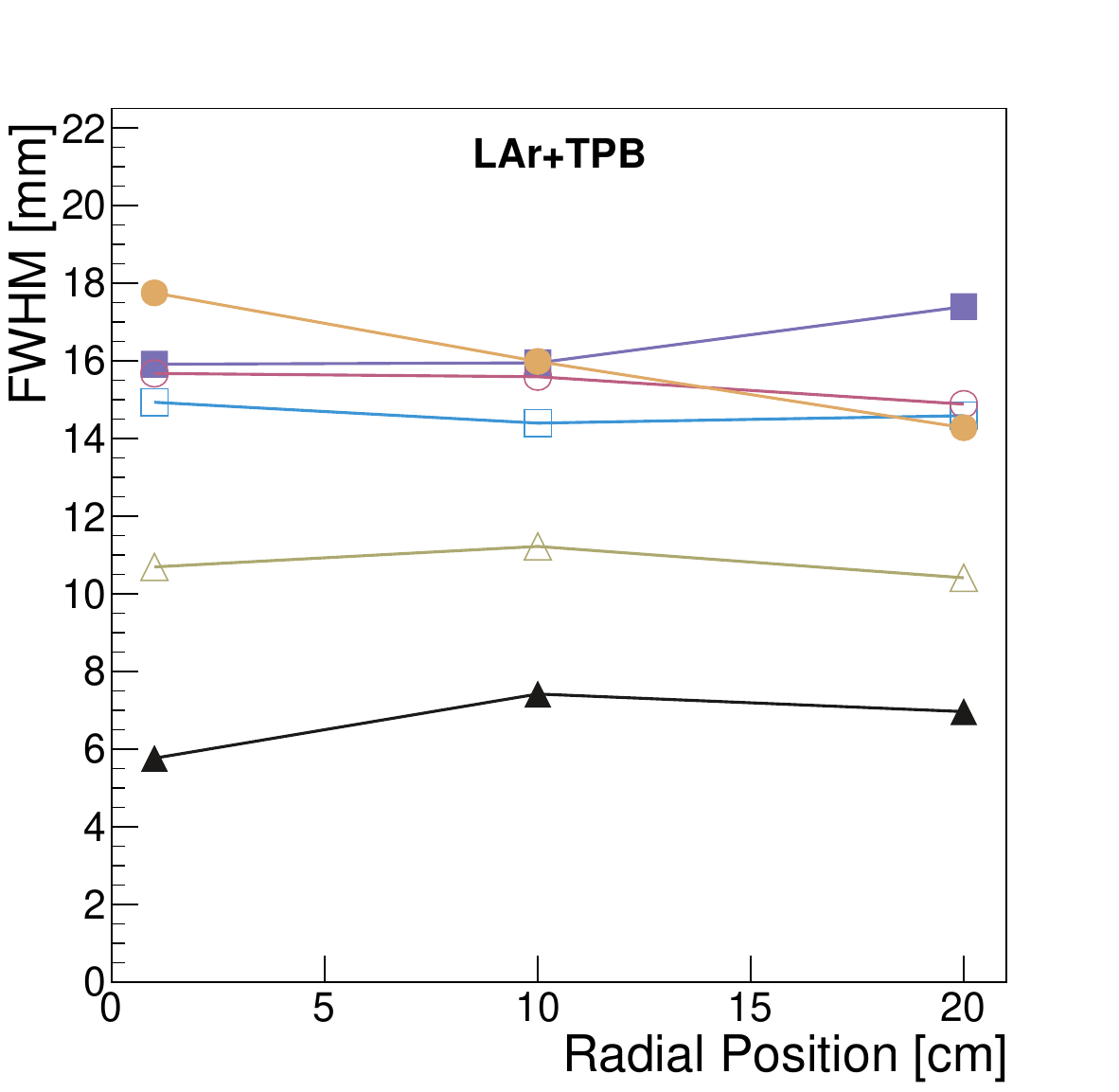}
\end{center}
\caption{The spatial resolution of reconstructed annihilation vertices was estimated without FDR. Radial, tangential, and axial resolutions FWHM for each radial position (1, 10, and 20 cm); Z-Center: at the center of the AFOV; Z-Offset: one-eighth of the AFOV along the axial axis from the end of the tomography device, before image reconstruction (raw data); {\bf Left:} Configuration LAr+Xe, {\bf Right:} Configuration LAr+TPB. The spatial resolution with LAr+Xe improves by more than a factor of \num{\sim 2} compared to resolutions with LAr+TPB, aligning with our expectations as discussed in the result section.}
\label{fig:TPB-Xe-FWHM}
\end{figure}

\begin{figure}[H]
\begin{center}
\includegraphics[width=0.9\columnwidth]{./legendborder.pdf}
\includegraphics[width=0.45\columnwidth]{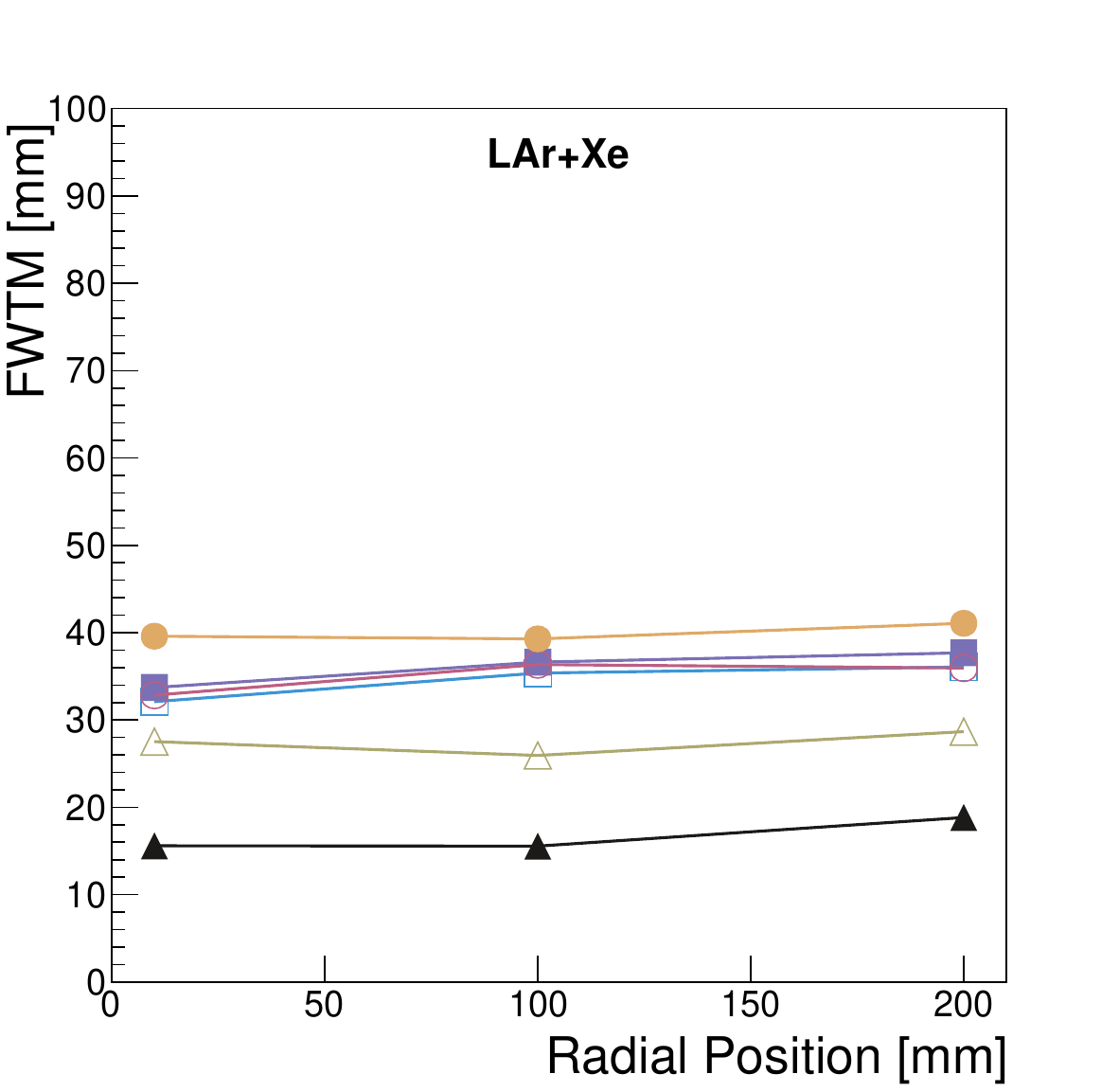}
\includegraphics[width=0.45\columnwidth]{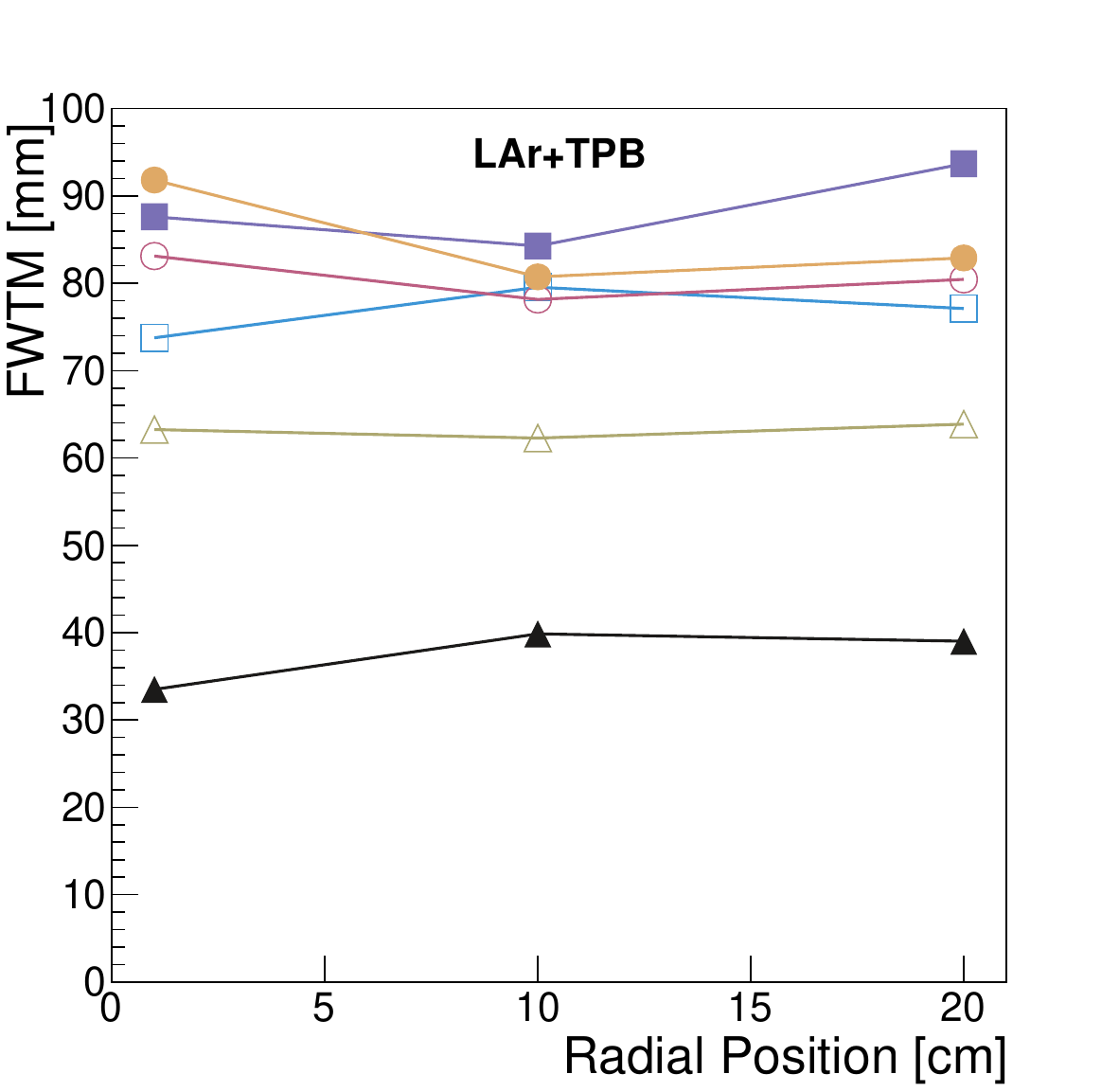}
\end{center}
\caption{Radial, tangential, and axial resolutions FWTM for each radial position (1, 10, and 20 cm), Z-Center: at the center of the AFOV;  Z-Offset: one-eighth of the AFOV along the axial axis from the end of the tomography device.{ \bf Left:} Configuration LAr+Xe, { \bf Right:} Configuration LAr+TPB.}
\label{fig:TPB-Xe-FWTM}
\end{figure}

\begin{figure}[H]
\begin{center}
\includegraphics[width=\columnwidth]{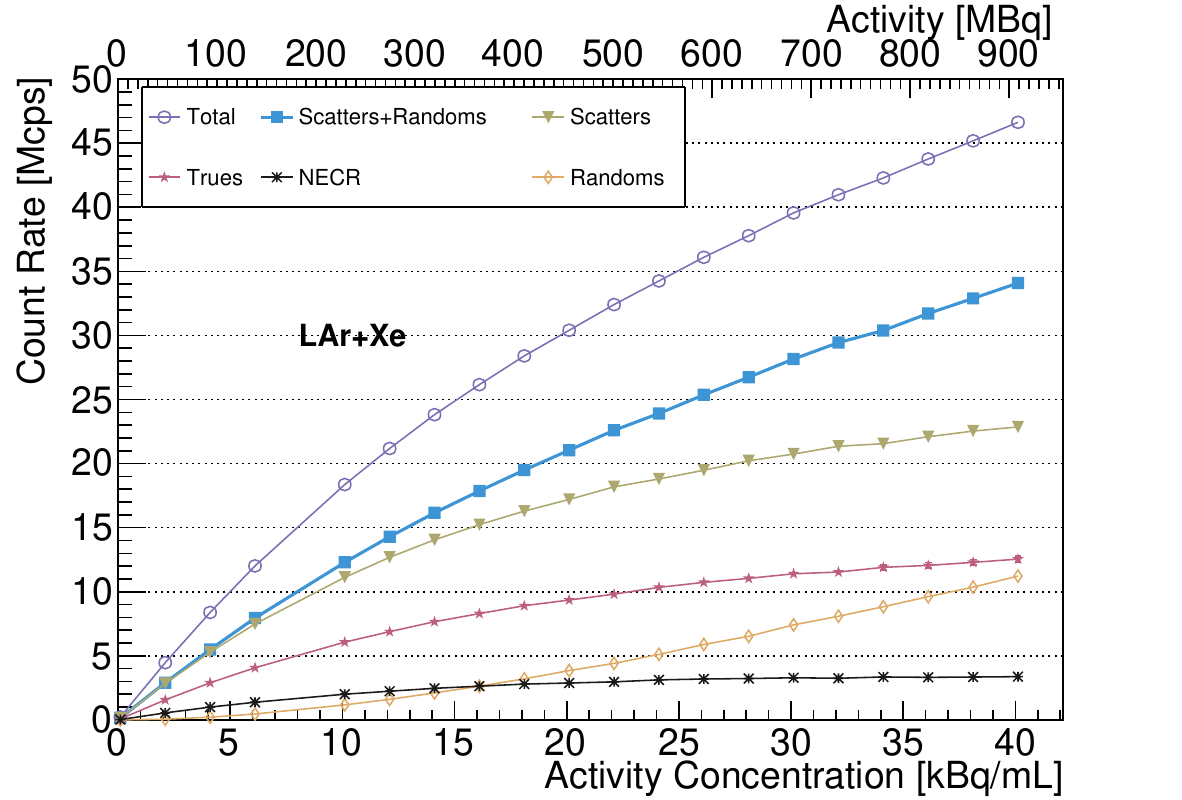}
\end{center}
\caption{Count rates without OPC, LAr+Xe configuration.}
\label{fig:xe-cut0}
\end{figure}

\begin{figure}[H]
\begin{center}
\includegraphics[width=\columnwidth]{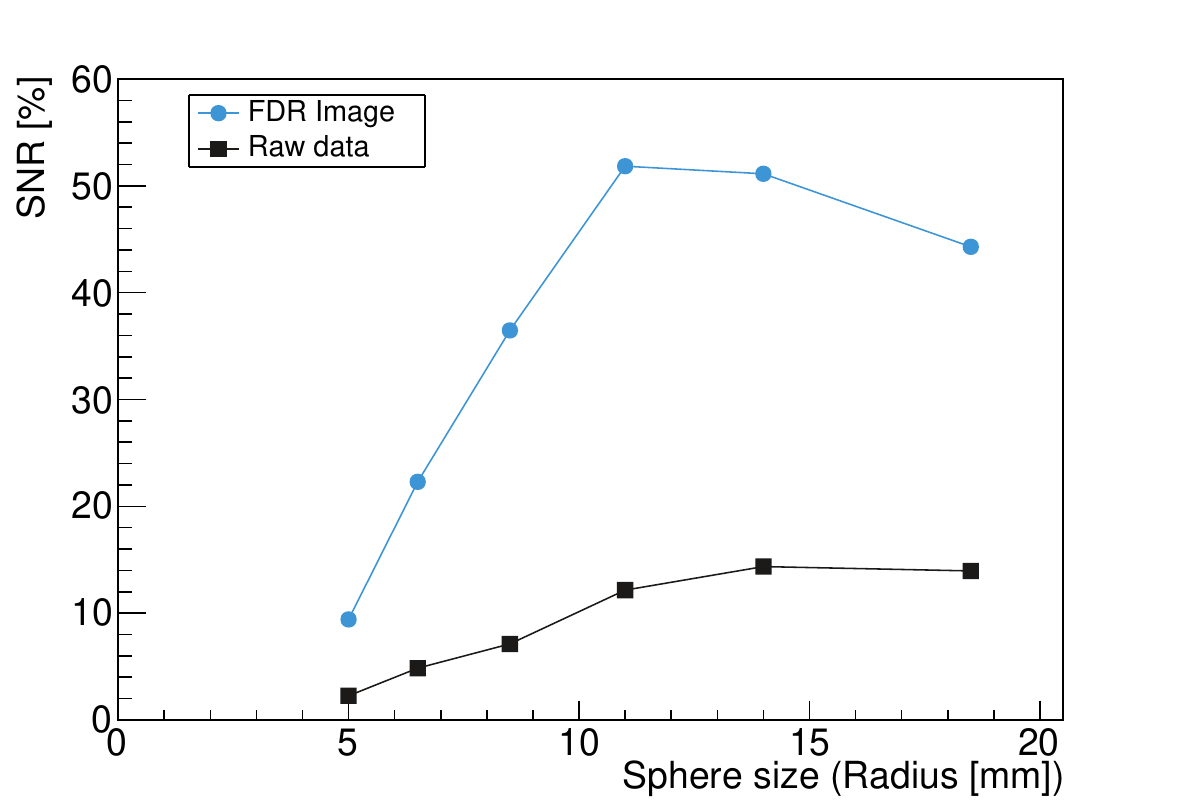}
\end{center}
\caption{Comparison of SNR between FDR images and raw data.}
\label{fig:SNR}
\end{figure} 

\begin{figure}[H]
\begin{center}
\includegraphics[width=\columnwidth]{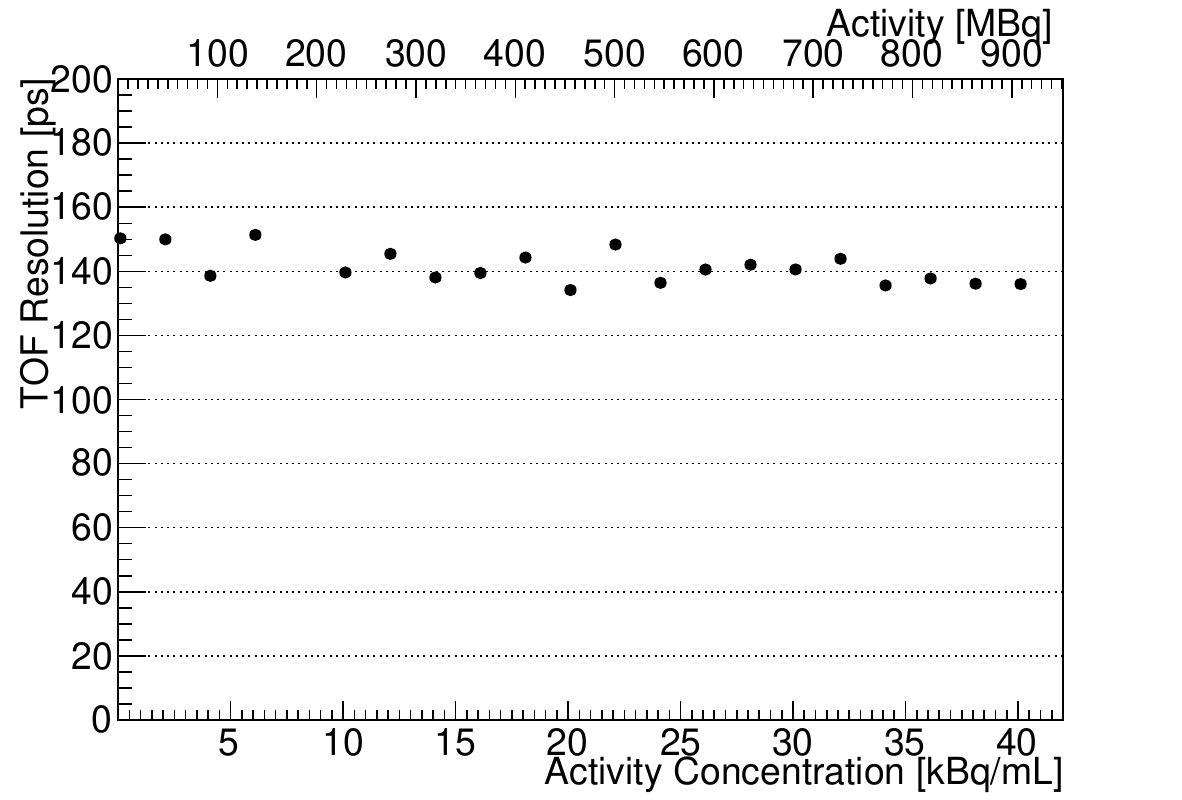}
\end{center}
\reducefigurespace
\caption{TOF resolution vs. activity concentration.}
\label{fig:TOF}
\end{figure}

\begin{equation}
\label{eq:PC}
PC_j = \frac{\frac{C_{H,j}}{C_{B,j}}-1}{\frac{ac_{H,j}}{ac_{B,j}}-1}\times 100 \%.\\
\end{equation}
where
\begin{itemize}
\item $C_{H,j}$ is the average counts in the regions of interest (ROI) for sphere j.
\item $B_{H,j}$ is the average of the background ROI counts for sphere j.
\item $ac_{H,j}$ is the activity concentration in the hot spheres;
\item $ac_{B,j}$ is the activity concentration in the background;
\end{itemize}

\begin{equation}
\label{eq:SD}
SD_j=\sqrt{\frac{\sum_{k=1}^{K}(C_{B,k,j}-C_{B,j})^2}{K-1}}.
\end{equation}
where $SD_j$ is the standard deviation of the background ROI counts for sphere j and the sum is taken over the K=60 background ROIs.
\begin{equation}
\label{eq:BV}
BV_j = \frac{SD_j}{C_{B,j}}\times 100 \%.
\end{equation}

\begin{equation}
\label{eq:RE}
RE = \frac{C_{lung}}{C_{B,37mm}}\times 100 \%.
\end{equation}
where 
\begin{itemize}
\item $C_{lung}$ is the average counts in the lung insert ROI;
\item $C_{C_{B,37mm}}$ is the average of the sixty 37 mm background ROIs;
\end{itemize}

\begin{equation}
SNR_j= \frac{C_{H,j}-C_{B,j}}{SD_j}\times 100 \%.
\label{eq:SNReq}
\end{equation}

\end{document}